\documentclass[allclo]{FBSart}
\usepackage{amsfonts}
\usepackage{amssymb}
\usepackage{graphicx} 

\newcommand{\lsim}{\mathrel{\rlap{\lower4pt\hbox{\hskip0pt$\sim$}} 
\raise1pt\hbox{$<$}}}           
\newcommand{\gsim}{\mathrel{\rlap{\lower4pt\hbox{\hskip0pt$\sim$}} 
\raise1pt\hbox{$>$}}}           

\usepackage[T1]{fontenc}
\usepackage[latin1]{inputenc}

\providecommand{\tabularnewline}{\\}

\newcommand{\Eq}{Eq.\,}

\usepackage{color}
\definecolor{purple}{rgb}{0.5,0,0.5}

\usepackage[colorlinks=true,
            pdfstartview=FitV,
        linkcolor=purple,
        citecolor=purple,
        urlcolor=blue,
        pdftitle={Lattice Variational Approach},
        pdfauthor={Stewart V. Wright},
        pdfpagemode=None
       ]{hyperref}

\begin{document}

\title{Schwinger functions and light-quark bound states}

\author{M.\,S.\ Bhagwat,\instnr{1} A.\ H\"oll,\instnr{2} A.\ Krassnigg,\instnr{3}
C.\,D.\ Roberts\instnr{1} and S.\,V.\ Wright\instnr{1}} 

\instlist{Physics Division, Argonne National Laboratory, 
             Argonne, IL 60439-4843, U.S.A. \and
Institut f\"ur Physik, Universit\"at Rostock, D-18051 Rostock, Germany \and 
Institut f\"ur Physik, 
        Karl-Franzens-Universit\"at Graz, A-8010 Graz, Austria
}  

\runningauthor{M.\,S.\ Bhagwat,  A.\ H\"oll, A.\ Krassnigg, 
C.\,D.\ Roberts and S.\,V.\ Wright}
\runningtitle{Schwinger functions and light-quark bound states}
\sloppy

\maketitle        

\begin{abstract}
We examine the applicability and viability of methods to obtain knowledge about bound-states from information provided solely in Euclidean space.  Rudimentary methods can be adequate if one only requires information about the ground and first excited state and assumptions made about analytic properties are valid.  However, to obtain information from Schwinger functions about higher mass states, something more sophisticated is necessary.  A method based on the correlator matrix can be dependable when operators are carefully tuned and errors are small.  This method is nevertheless not competitive when an unambiguous analytic continuation of even a single Schwinger function to complex momenta is available.
\end{abstract}

\section{Introduction}
The probability measure plays a crucial role in quantum field theory.  The simplest such measure is the Gaussian distribution
\begin{equation}
 {\cal K}_\tau(q,q^\prime)= 
\frac{1}{(2\pi \tau)^{3/2}} \,\exp\left[-\frac{\left(q-q^\prime\right)^2}{2
\tau}\right]\,,
\end{equation}
which is the fundamental solution of the heat equation and the probability density that indicates whether a particle characterised initially by a coordinate $q$ will be described by $q^\prime$ after an interval $\tau$.  This distribution, which is related to the free-particle density matrix in statistical mechanics, is a probability measure because it is positive definite and normalisable [actually, normalised: $\lim_{\tau\to 0}{\cal K}_\tau(q,q^\prime) = \delta^3(q-q^\prime)$].  In Euclidean space a theory's generating functional can truly be expressed in terms of a probability measure and the properties of such measures make it likely that the rigorous definition of interacting quantum field theories is possible in that case.  However, in Minkowski space the probability density becomes a probability \emph{amplitude}, through the appearance of ``$i$'' in the exponent, and that precludes the formulation of a measure theory.

The moments of a probability measure are $n$-point Schwinger functions.  They correspond to vacuum expectation values of Euclidean fields and may loosely be termed Euclidean space Green functions.  When certain conditions are met \cite{glimm}, analytic continuation of the Schwinger functions yields the Wightman functions and one may prove the reconstruction theorem; namely, the complete content of a quantum field theory is recovered from the Wightman functions.\footnote{NB.\ Minkowski space Green functions are constructed from appropriately time-ordered combinations of Wightman functions.}  This is the basis for the contemporary belief that the evaluation of a theory's Schwinger functions is equivalent to solving that theory.

While that may be true in principle one must nonetheless develop practical means by which to extract physical information from the Schwinger functions.  The challenge may be illustrated by noting that complete knowledge of a two-point Schwinger function in momentum space corresponds only to direct knowledge of the expectation value at spacelike momenta.  A physical particle pole can only appear at timelike momentum.  Locating such a pole therefore requires that all the conditions be met for a unique analytic continuation.  If the Schwinger function is known only at a discrete set of points; i.e., on a set of measure zero, then that is strictly impossible.  In this common instance all that is really possible is constrained inference and an estimation of the inherent error.

For us a context for these observations is the problem of determining the spectrum of QCD.  Much has been learnt about the two-point Schwinger functions for quarks, gluons and ghosts.  A consensus has been reached; viz., these functions cannot be described by a positive definite spectral density.  However, the pointwise behaviour of the continuation of these functions to timelike momenta is uncertain -- see, e.g., Refs.\,\cite{hawes,mandarquark,fischergluon,detmoldquark,fischerquark,sauli} and references therein --  and is the subject of continuing study, e.g., Ref.\,\cite{bhagwatpanic,nickel}. 

These two-point functions appear in the kernels of the Bethe-Salpeter equations whose solutions provide information about the properties of colour-singlet hadrons.  The theory and phenomenology of such applications are reviewed in Refs.\,\cite{cdragw,Roberts:2000aa,ralvs,pmcdr,cf06}.  The study of ground state mesons in the pseudoscalar and vector channels has met with success, as evidenced by recent applications to heavy-heavy mesons \cite{heavyheavy} and vector meson electromagnetic form factors \cite{bhagwatmaris}.  A present challenge is the extension to excited states in these and other channels, e.g., Refs.\,\cite{Holl:2004fr,arnebeijing,andreasradialgg,Krassnigg:2006ps}, and to channels in which the ground state lies above $1\,$GeV, e.g., Refs.\,\cite{a1b1,burdenexotic,watsona1,krassnigga1}.\footnote{There is also merit in studying the colour antitriplet quark-quark channel since such diquark correlations quite likely play an important role in baryon structure, e.g, \protect\cite{arneN,arneN2,Flambaum:2005kc,arneN3}.}  A veracious analytic continuation of two-point Schwinger functions becomes increasingly important as the bound-state mass becomes larger \cite{bhagwatpoles}.  The possibility that this continuation might be absent returns us to the question: can reliable information about bound state properties be obtained directly from the Schwinger functions?

Another context for this question is the numerical simulation of lattice-regularised QCD.  That approach is grounded on the Euclidean space functional integral.  Schwinger functions are all that it can directly provide.  Hence it can only be useful if methods are found so that the question can be answered in the affirmative.  It is therefore unsurprising that lattice-QCD practitioners have expended much effort on this problem (see, e.g., Refs.\,\cite{latticemem,Leinweber:2004it} and references therein).  Our study explores the efficacy of the methods devised for lattice-QCD and whether it is worthwhile to adapt them to continuum approaches.

In Sect.\,\ref{sec:Inhomogeneous-Dyson--Schwinger-Equation} we recapitulate on aspects of Dyson-Schwinger equations (DSEs) that are relevant to the study of mesons, focusing in particular on the inhomogeneous Bethe-Salpeter equation for colour-singlet three-point Schwinger functions.  In Sect.\,\ref{sec:Methods} we explore and illustrate the application of a simple momentum-space method to the problem of extracting meson masses and vertex residues from these Schwinger functions.  This is followed by an examination of two straightforward configuration-space methods for recovering bound-state information from Schwinger functions, and a study of the impact of statistical noise on these procedures.  The picture that emerges is that rudimentary methods are often adequate if one only requires information about the ground and first excited state, and assumptions made about the Schwinger function's analytic structure are valid.  However, to obtain information from Schwinger functions about heavier states in a given channel, something much more refined is necessary.  In Sect.\,\ref{correlatormatrixmethod} we analyse one such method; namely, that based on the correlator matrix \cite{Michael:1985ne,Luscher:1990ck}.  Section~\ref{epilogue} is an epilogue.

\section{Inhomogeneous Bethe--Salpeter Equation}
\label{sec:Inhomogeneous-Dyson--Schwinger-Equation}
A calculation of the properties of two-body bound states in quantum field theory may begin with the Poincar\'{e} covariant Bethe-Salpeter equation (BSE). The inhomogeneous BSE for a pseudoscalar quark-antiquark vertex is\footnote{We work with two degenerate quark flavours and hence Pauli matrices are sufficient to represent the flavour structure.  In our Euclidean metric:  $\{\gamma_\mu,\gamma_\nu\} = 2\delta_{\mu\nu}$; $\gamma_\mu^\dagger = \gamma_\mu$; and $a \cdot b = \sum_{i=1}^4 a_i b_i$.  For a timelike vector $P_\mu$, $P^2<0$.  More about the metric conventions can be found in Sect.\,2.1 of Ref.\,\protect\cite{Roberts:2000aa}.}
\begin{equation}
\left[\Gamma_{5}^{j}(k;P)\right]_{tu}=Z_{4}\gamma_{5}\,\frac{\tau^{j}}{2}
 + \int_{q}^{\Lambda}\left[\chi_{5}^{j}(q;P)\right]_{sr}K_{rs}^{tu}(q,k;P)\,,
\label{eq:DSE_inhomogeneous}
\end{equation}
where $\Gamma_{5}^{j}(k;P)$ is the vertex Bethe-Salpeter amplitude, with $k$ the relative and $P$ the total momentum; $r,\ldots,u$ represent colour, Dirac and flavour matrix indices;
\begin{equation}
\label{chidef}
\chi_{5}^{j}(q;P)=S(q_{+})\Gamma_{5}^{j}(q;P)S(q_{-})\,,
\end{equation}
$q_{\pm}=q\pm P/2$; $Z_4$ is the Lagrangian-mass renormalisation constant, described in connection with Eq.\,(\ref{Z4def}); and $\int_{q}^{\Lambda}$ represents a Poincar\'{e} invariant regularisation of the integral, with $\Lambda$ the regularisation mass-scale \cite{Maris:1997hd,Maris:1997tm}.   In Eq.\,(\ref{eq:DSE_inhomogeneous}), $K$ is the fully amputated and renormalised dressed-quark-antiquark scattering kernel and $S$ in Eq.\,(\ref{chidef}) is the renormalised dressed-quark propagator.  It is notable that the product $SSK$ is a renormalisation group invariant.  

The solution of Eq.\,(\ref{eq:DSE_inhomogeneous}) has the form 
\begin{eqnarray}
\nonumber
\lefteqn{i\Gamma_{5}^{j}(k;P) = \frac{\tau^{j}}{2}\gamma_{5}\left[iE_{5}(k;P)\right.  }\\
&& + \left. \gamma\cdot P\, F_{5}(k;P) + \gamma\cdot k\, k\cdot P\, G_{5}(k;P)+\sigma_{\mu\nu}\, k_{\mu}P_{\nu}\, H_{5}(k;P)\right]. 
\label{genpvv}
\end{eqnarray}
This is the structure necessary and sufficient to ensure Poincar\'{e} covariance.  It follows that quark orbital angular momentum is generally present within pseudoscalar and indeed all bound states.  This is quantified, e.g., in Ref.\,\cite{heavyheavy}.

The dressed-quark propagator appearing in the BSE's kernel is determined by the renormalised gap equation 
\begin{eqnarray}
S(p)^{-1} & =&  Z_2 \,(i\gamma\cdot p + m^{\rm bm}) + \Sigma(p)\,, \label{gendse} \\
\Sigma(p) & = & Z_1 \int^\Lambda_q\! g^2 D_{\mu\nu}(p-q) \frac{\lambda^a}{2}\gamma_\mu S(q) \Gamma^a_\nu(q,p) , \label{gensigma}
\end{eqnarray}
where $D_{\mu\nu}$ is the dressed-gluon propagator, $\Gamma_\nu(q,p)$ is the dressed-quark-gluon vertex, and $m^{\rm bm}$ is the $\Lambda$-dependent current-quark bare mass.  The quark-gluon-vertex and quark wave function renormalisation constants, $Z_{1,2}(\zeta^2,\Lambda^2)$ respectively, depend on the renormalisation point, $\zeta$, the regularisation mass-scale and the gauge parameter.  

The gap equation's solution has the form 
\begin{eqnarray} 
 S(p)^{-1} 
%
& =& \frac{1}{Z(p^2,\zeta^2)}\left[ i\gamma\cdot p + M(p^2)\right] . 
\label{sinvp} 
\end{eqnarray} 
It is obtained from Eq.\,(\ref{gendse}) augmented by the renormalisation condition
\begin{equation}
\label{renormS} \left.S(p)^{-1}\right|_{p^2=\zeta^2>0} = i\gamma\cdot p +
m(\zeta)\,,
\end{equation}
where $m(\zeta)$ is the renormalised (running) mass: 
\begin{equation}
\label{Z4def}
Z_2(\zeta^2,\Lambda^2) \, m^{\rm bm}(\Lambda) = Z_4(\zeta^2,\Lambda^2) \, m(\zeta)\,.
\end{equation}
Features of the gap equation and its solutions bear upon the radius of convergence for a perturbative expansion in the current-quark mass of physical quantities \cite{changlei}.

\begin{figure}[t]
\centerline{\includegraphics[%
  clip,
  width=0.80\textwidth]{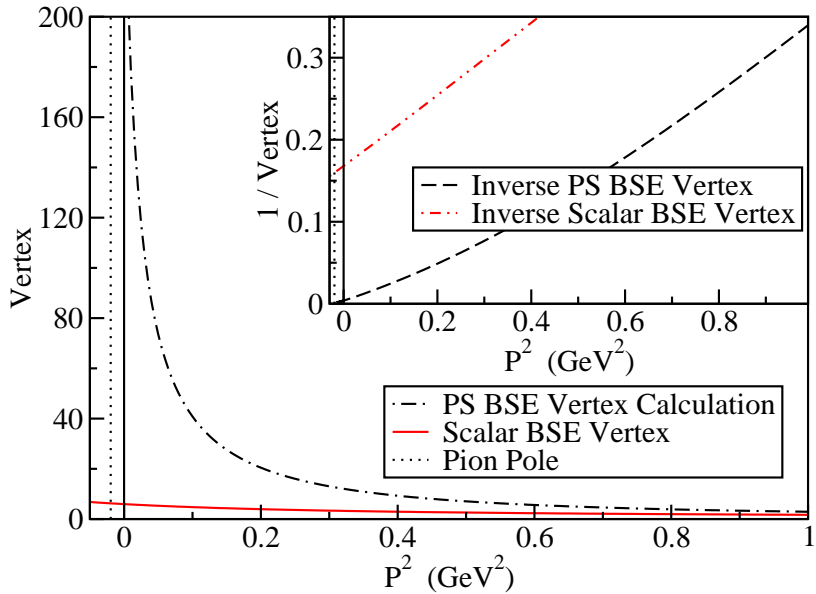}}
\caption{\label{fig:BSE_vertex} Dimensionless pseudoscalar (PS) and scalar amplitudes, $E_{5}(0;P^{2})$ and $E_{S}(0;P^{2})$, respectively, obtained by solving an inhomogeneous BSE of the type in Eq.\,(\protect\ref{eq:DSE_inhomogeneous}) using the renormalisation-group-improved rainbow-ladder truncation introduced in Ref.\,\protect\cite{maristandy1}.  (We set the model's mass-scale $\omega=0.33\,$GeV and used a current-quark mass $m_{u,d}(\zeta_{19})=3.7\,$MeV, $\zeta_{19}=19\,$GeV.  NB.\ For $\omega\in[0.3,0.5]\,$GeV, ground state observables are constant along a trajectory $\omega D = (0.72 \, {\rm GeV})^3 =: m_g^3$ \protect\cite{raya}.)  The vertical dotted-line indicates the position of the ground state $\pi$ mass pole; whereas the vertical solid line marks only the coordinate zero.  The inset shows $1/E_{5}(0;P^{2})$ and $1/E_{S}(0;P^{2})$.}
\end{figure}

The solution of the inhomogeneous equation, Eq.\,(\ref{eq:DSE_inhomogeneous}), exists for all values of $P^2$, timelike and spacelike, with each bound state exhibited as a pole.  This is illustrated in Fig.\,\ref{fig:BSE_vertex}, wherein the solution is seen to evolve smoothly with $P^{2}$ and the pole associated with the pseudoscalar ground state is abundantly clear.\footnote{NB.\ The kernel used for this calculation \protect\cite{maristandy1} has hitherto provided only stable bound-states.  For a resonance the pole is not located on the real axis.  Nevertheless, in principle such systems can also be handled via the BSE.  Herein we will mainly overlook such cases because the associated complications are not yet relevant: all contemporary studies employ kernels that omit the channels associated with physical decays; and unstable states are still not much studied using lattice-QCD.}  

Naturally, the numerical determination of the precise location of the first pole (ground state) in $E_5(k^2=0;P^2)$ will generally be difficult.  The task becomes harder if one seeks in addition to obtain the positions of excited states.  It is for these reasons that the homogeneous equation \cite{llewellyn} is usually used.  The homogeneous pseudoscalar BSE is obtained from Eq.\,(\ref{eq:DSE_inhomogeneous}) by omitting the driving term; viz., the matrix valued constant $(1/2) Z_{4}\gamma_{5} \tau^{j}$.  The equation thus obtained defines an eigenvalue problem, with the bound state's mass-squared being the eigenvalue and its Bethe-Salpeter amplitude, the eigenvector.  As such, the homogeneous equation only has solutions at isolated timelike values of $P^2$.  However, if one is employing a framework that can only provide reliable information about the form of the Bethe-Salpeter vertex at spacelike momenta; i.e., a framework that only provides Schwinger functions, then a scheme must be devised that will yield the pole positions from this information alone.

\section{Masses from Schwinger Functions: Simple Methods}
\label{sec:Methods}
\subsection{Momentum space: Pad\'e approximant}
\label{mompade}
One straightforward approach is to focus on 
\begin{equation}
P_E(P^2)=\frac{1}{E(k^2=0;P^2)}
\end{equation}
and locate its zeros.  It is plain from Fig.\,\ref{fig:BSE_vertex} that this method can at least be successful for the ground state in the pseudoscalar channel.  It is important to determine whether the approach is also practical for the determination of some properties of excited states.

For the purpose of exploration and illustration, consider a model for an inhomogeneous vertex (three-point Schwinger function) whose analytic structure is known precisely; namely,
\begin{equation}
V(P^{2})=b+\sum_{i=0}^{M-1}\frac{a_{i}}{P^{2}+m_{i}^{2}}\,,
\label{eq:simple_model}
\end{equation}
where: for each $i$, $m_{i}$ is the bound state's mass and $a_{i}$ is the residue of the bound state pole in the vertex, which is related to the state's decay constant (see \ref{App:residue}); and $b$ is a constant that represents the perturbative background that is necessarily dominant at ultraviolet total momenta.  The particular parameter values we employ are listed in Table\,\ref{tab:Model_Param}.   This \textit{Ansatz} provides an information sample that expresses essential qualitative features of true DSE solutions for colour-singlet three-point Schwinger functions.

\begin{table}[t]
\caption{\label{tab:Model_Param} Parameters characterising the vertex \emph{Ansatz},
Eq.\,(\protect\ref{eq:simple_model}). They were chosen without prejudice, subject to the constraint in quantum field theory that residues of poles in an observable projection of a three-point Schwinger function should alternate in sign \cite{Holl:2004fr}, and ordered such that $m_{i}<m_{i+1}$, with $i=0$ denoting the ground state and $i=1$ the first excited state, etc.  $b=0.78$ is the calculated value of $Z_{4}(\zeta_{19},\Lambda=200\,\textrm{GeV})$ used to obtain the curves in Fig.\,\protect\ref{fig:BSE_vertex}.}
\begin{center}
\begin{tabular*}{0.50\textwidth}{
|c@{\extracolsep{0ptplus1fil}}
|c@{\extracolsep{0ptplus1fil}}|c@{\extracolsep{0ptplus1fil}}|}\hline
$i$~ & Mass, $m_i$ (GeV) & Residue, $a_i$ (GeV$^2$)\\\hline
0& 0.14& ~4.23\\  
1& 1.06& -5.6~ \\
2&1.72& ~3.82\\  
3&2.05 & -3.45 \\  
4&2.2~ & ~2.8~ \\\hline
\end{tabular*}
\end{center}
\end{table}

To proceed we employ a diagonal Pad\'e approximant of order $N$ to analyse the information sample generated by Eq.\,(\ref{eq:simple_model}); viz., 
\begin{equation}
f_{N}(x) = \frac{c_{0}+c_{1}x+\ldots+c_{N}x^N} {1+c_{N+1} x+\ldots+c_{2N}x^N}\,, \; x = P^2\,,
\label{eq:pade}
\end{equation}
is used to fit a sample of values of $1/V(P^2)$ defined on a discrete $P^2$ grid, such as would be employed in a numerical solution of the BSE.  The known ultraviolet behaviour of typical vertices requires a diagonal approximant.\footnote{NB.\ A real-world data sample will exhibit logarithmic evolution beyond the renormalisation point.  No simple Pad\'e approximant can recover that.  However, this is not a problem in practical applications because the approximant is never applied on that domain.}  There are some similarities between this approach and that of the inverse amplitudes method in Refs.\,\cite{Truong:1988zp,Dobado:1992ha}.

In a confining theory it is likely that a colour-singlet three-point function exhibits a countable infinity of bound state poles.  Therefore no finite order approximant can be expected to recover all the information contained in that function.  The vertex model presented in Eq.\,(\ref{eq:simple_model}) possesses $M$ bound state poles.  It is reasonable to expect that an approximant of order $N<M$ can at most provide reliable information about the first $N-1$ bound states, with the position and residue associated with the $N^{\rm th}$ pole providing impure information that represents a mixture of the remaining $M-(N-1)$ signals and the continuum.  We anticipate that this is the pattern of behaviour that will be observed with any rank-$N$ approximation to a Schwinger function.  The $N$-dependence of the Pad\'e fit can provide information on this aspect of the procedure.  The domain of spacelike momenta for which information is available may also affect the reliability of bound-state parameters extracted via the fitting procedure.  Information on this possibility is provided by fitting Eq.\,(\ref{eq:pade}) to the \textit{Ansatz} data on a domain $[0,P_{\mathrm{max}}^{2}]$, and studying the $P_{\mathrm{max}}^{2}$-dependence of the fit parameters.

\begin{figure}[t]
\centerline{\includegraphics[clip,width=0.80\textwidth]{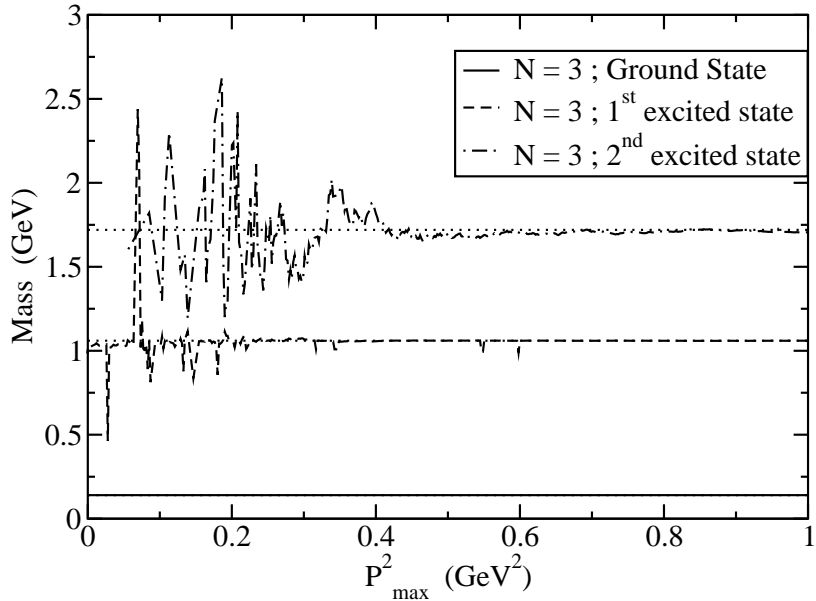}}

\caption{\label{fig:3pade_mass} Pole positions (mass values) obtained through a fit of Eq.\,(\protect\ref{eq:pade}) with $N=3$ to data for $1/V(P^2)$ generated from Eq.\,(\protect\ref{eq:simple_model}) with the parameters listed in Table \ref{tab:Model_Param}.  The coordinate $P_{\mathrm{max}}^{2}$ is described in the text. Horizontal dotted lines indicate the three lightest masses in Table \ref{tab:Model_Param}.  The ground state mass (solid line) obtained from the Pad\'e approximant lies exactly on top of the dotted line representing the true value.\vspace*{-4ex}}
\end{figure}

\begin{figure}[t]
\centerline{\includegraphics[clip,width=0.80\textwidth]{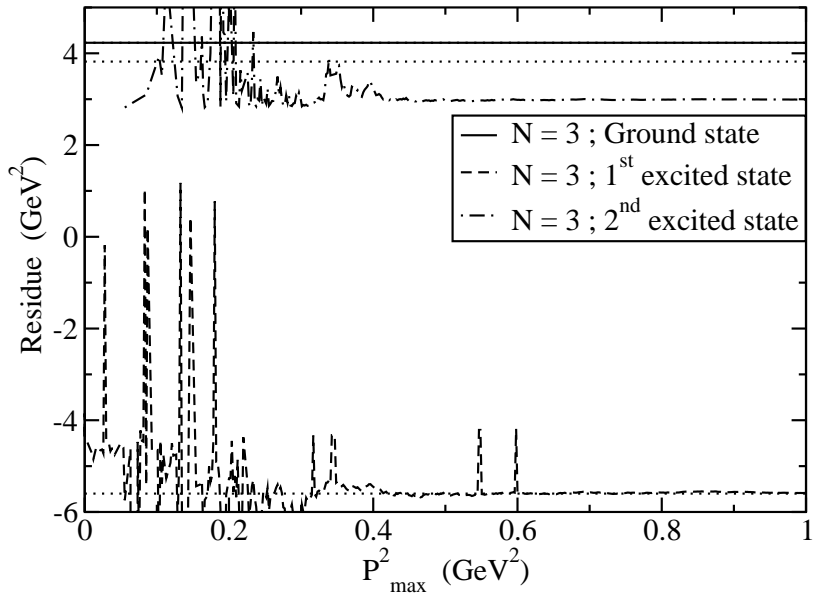}}

\caption{\label{fig:3pade_residue} Pole residues obtained through a fit of Eq.\,(\protect\ref{eq:pade}) with $N=3$ to data for $1/V(P^2)$ generated from Eq.\,(\protect\ref{eq:simple_model}) with the parameters in Table \ref{tab:Model_Param}.  Horizontal dotted lines indicate the residues associated with the three lightest masses in Table \ref{tab:Model_Param}.  The residue associated with the ground state (solid line) lies exactly atop the dotted line representing the true value.  The residue for the second pole exhibits a plateau at the correct (negative) value.  However, the plateau exhibited by the result for the residue of the third pole is wrong.\vspace*{-4ex}}
\end{figure}

We find that a Pad\'e approximant fitted to $1/V(P^2)$ can accurately recover the pole residues and locations associated with the ground and first excited states.  This is illustrated in Fig.\,\ref{fig:3pade_mass}, which exhibits the $P_{\mathrm{max}}^{2}$-dependence of the mass-parameters determined via a $N=3$ Pad\'e approximant.  Plateaux appear for three isolated zeros, which is the maximum number possible, and the masses defined by these zeros agree very well with the three lightest values in Table \ref{tab:Model_Param}.  This appears to suggest that the procedure has performed better than anticipated.  However, that inference is seen to be false in Fig.\,\ref{fig:3pade_residue}, which depicts the $P_{\mathrm{max}}^{2}$-dependence of the pole residues.  While the results for $a_{0,1}$ are correct, the result inferred from the plateau for $a_2$ is incorrect.  Plainly, if the value of $a_2$ were not known \textit{a priori}, then one would likely have been misled by the appearance of a plateau and produced an erroneous prediction from the fit to numerical data.  A $N=3$ approximant can truly at most only provide reliable information for the first $N-1=2$ bound states. 

\subsection{A realistic test}
\begin{figure}[t]
\begin{center}
\centerline{\includegraphics[%
  clip,
  width=0.36\textwidth,angle=-90]{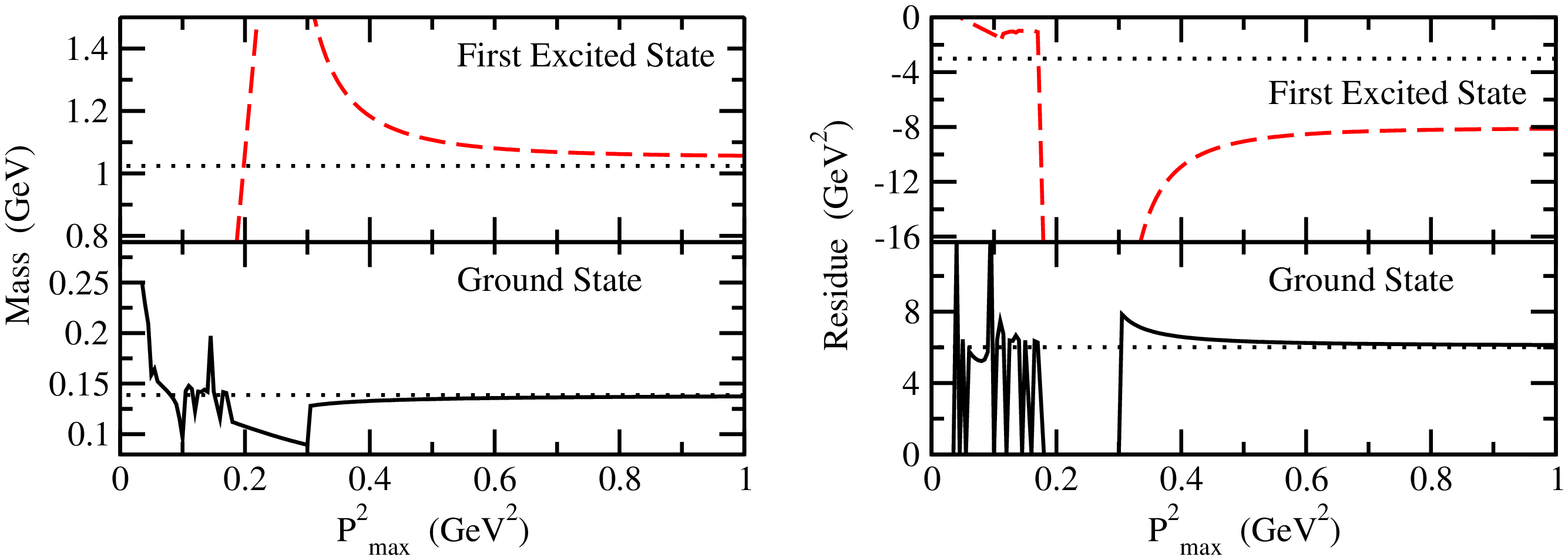}}

\parbox{\textwidth}{\caption{\label{fig:ps_meson_mass_residue} Mass and residue obtained from a fit of \Eq(\ref{eq:pade}) with $N=3$ to the DSE result for $1/E_{5}(k^{2}=0;P^{2})$ in Fig.\,\protect\ref{fig:BSE_vertex}.  Horizontal dotted lines indicate the masses and residues obtained for the ground and first excited state via a direct solution of the homogeneous BSE.}}

\end{center}

\begin{center}
\centerline{\includegraphics[%
  clip,
  width=0.39\textwidth,angle=-90]{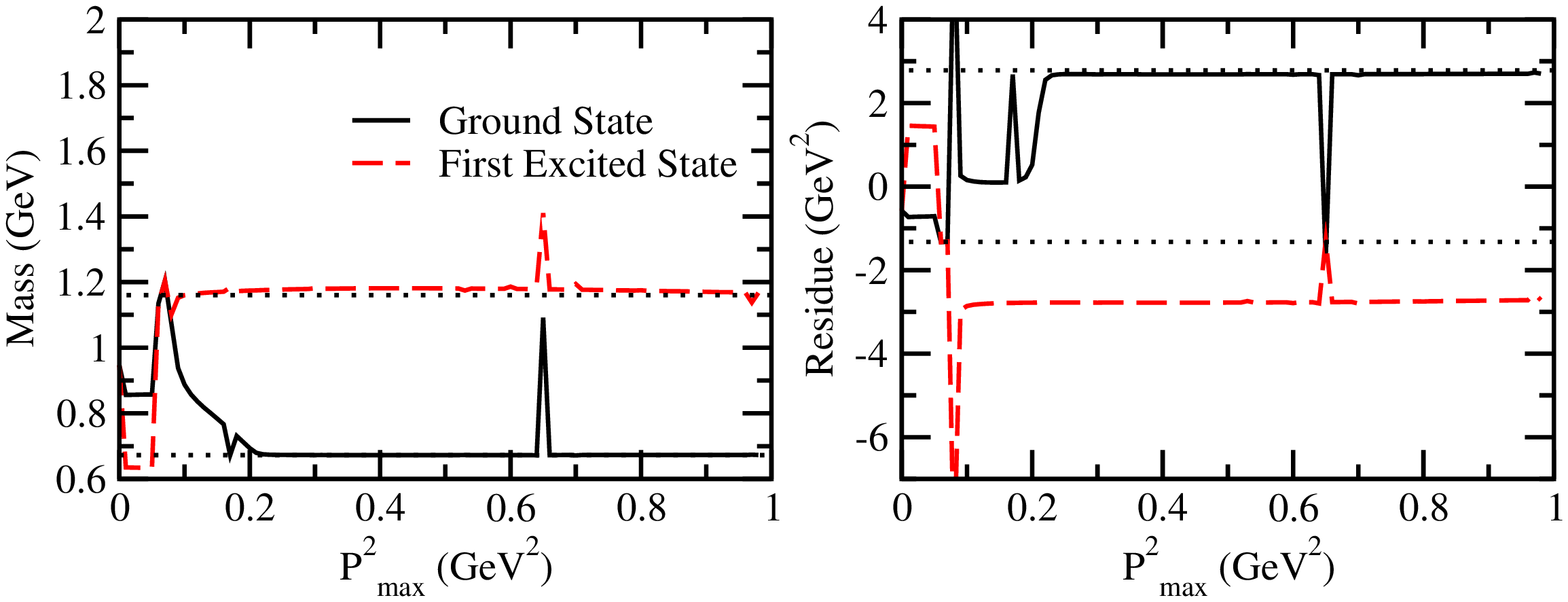}}

\parbox{\textwidth}{\caption{\label{fig:scalar} Mass and residue obtained from a fit of \Eq(\ref{eq:pade}) with $N=3$ to the DSE result for $1/E_{S}(k^{2}=0;P^{2})$; i.e., for the scalar channel, also depicted in Fig.\,\protect\ref{fig:BSE_vertex}.  As in Fig.\,\protect\ref{fig:ps_meson_mass_residue}, horizontal dotted lines indicate the masses and residues obtained for the ground and first excited state via a direct solution of the homogeneous BSE.}}

\end{center}
\end{figure}

Following this exploration of the viability \emph{in principle} of using spacelike data and the Pad\'e method to extract bound state information, we considered the DSE-calculated pseudoscalar vertex that is in depicted Fig.\,\ref{fig:BSE_vertex}.  The masses and residues (see the appendix) for the ground state pseudoscalar and its first radial excitation were obtained from the homogeneous BSE in Ref.\,\cite{Holl:2004fr}.  The comparison between these masses and those inferred from the Pad\'e approximant is presented in Fig.\,\ref{fig:ps_meson_mass_residue}.  It appears that with perfect (effectively noiseless) spacelike data at hand, reliable information on the masses can be obtained.  However, while the ground state residue is correctly determined, that of the first excited state is not.  We will subsequently explain this result.

In the rainbow-ladder truncation of QCD's DSEs the next heaviest ground-state meson, after the pseudoscalar, is the scalar \cite{Roberts:2000aa}.  The inhomogeneous scalar BSE vertex can be calculated \cite{Hoell:2005st} and the analysis described above repeated.  The relative magnitude of the scalar vertex compared with the pseudoscalar is shown in Fig.\,\ref{fig:BSE_vertex}.  In the vicinity of $P^2=0$ the signal for a bound state is much suppressed because in this model the mass-squared of the quark-core in the ground-state scalar meson is $P^2=-(0.675\,$GeV$)^2=-0.456\,{\rm GeV}^2$.  The contribution to the signal from a more massive state will be further suppressed.  
Figure \ref{fig:scalar} shows the results from an $N=3$ fit of Eq.\,(\ref{eq:pade}) to $1/E_{S}(k^{2}=0;P^{2})$ calculated using the inhomogeneous scalar BSE.  The two plateaux give masses in agreement with results obtained directly from the homogeneous BSE; i.e., $m_{1\,^3\!L_0} = 0.675\,$GeV and $m_{2\,^3\!L_0} = 1.16\,$GeV.  However, as with the pseudoscalar, the residue associated with the first excited state is wrong.  

\begin{figure}[t]
\centerline{\includegraphics[%
  clip,
  width=0.80\textwidth]{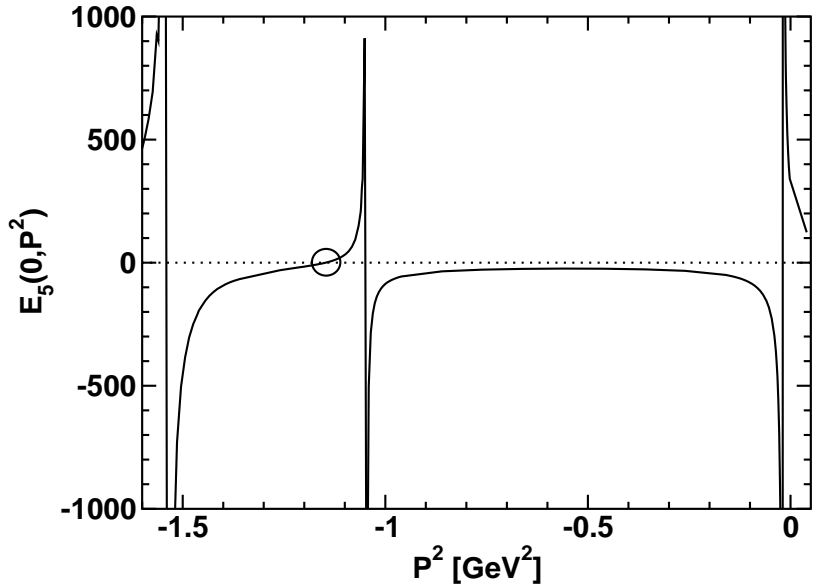}}

\parbox{\textwidth}{\caption{\label{timelike} Solution of Eq.\,(\ref{eq:DSE_inhomogeneous}) obtained directly at timelike momenta.  Three singularities that would normally indicate bound state poles are readily apparent.  However, the \emph{circle} at \mbox{$P^2=-1.15\,$GeV$^2$} marks an unanticipated zero.}}

\end{figure}

The pattern exposed here is not in accord with the expectations raised in Sect.\,\ref{mompade}.  However, this is only evident because we can simultaneously use timelike information to calculate the residues.  Success with the method explained in Sect.\,\ref{mompade} is predicated upon the assumption that the colour-singlet inhomogeneous vertex can be described by a positive definite spectral density; e.g., this validates the form of Eq.\,(\ref{eq:simple_model}).  The method may fail, therefore, if the interaction produces a vertex that is inconsistent with this physical requirement.  

To explore that possibility we employed matrix inversion methods to solve Eq.\,(\ref{eq:DSE_inhomogeneous}) directly for timelike total momenta.  The result, presented in Fig.\,\ref{timelike}, reveals a signal defect in the interaction model proposed in Ref.\,\cite{maristandy1}.  The model produces a vertex that possesses a \emph{zero} at $P^2=-1.15\,$GeV$^2$, and a zero in $E_5(0,P^2)$ is only possible if the spectral density in the colour-singlet pseudoscalar channel is negative on a measurable domain.  Similar behaviour is found in the scalar channel.

An explanation of Figs.\,\ref{fig:ps_meson_mass_residue} and \ref{fig:scalar} is now evident.  The zeros in $1/E_5(0,P^2)$ provide a strong signal and may readily be recovered via a Pad\'e approximant.  However, it is plain that the singularity in $1/E_5(0,P^2)$, produced by the zero circled in Fig.\,\ref{timelike} and lying just a little further into the timelike region than the position of the first excited state, can distort the value inferred for that state's residue.  

We have thus arrived via a circuitous route at important news.  In attempting to obtain knowledge of bound states from information provided solely at spacelike momenta, we have uncovered a weakness in the interaction model of Ref.\,\cite{maristandy1} that can plausibly be said to limit its domain of reliability to bound systems with mass $\lsim 1.1\,$GeV.  However, in order to establish this we had to forgo the notion of using spacelike information alone.  Without knowing the correct values of the residues \emph{a priori}, one may reasonably have accepted the results in Figs.\,\ref{fig:ps_meson_mass_residue} and \ref{fig:scalar} as true measures of the first excited states' properties.  The analysis in this section emphasises that in relying solely on spacelike information one rests heavily on assumptions made about the analytic properties of the Schwinger functions under examination.

\subsection{Configuration space analysis}
\label{sec:coord_space}
It is natural to ask whether more information about bound states can be extracted from Schwinger functions specified in configuration space.  These are, for example, the quantities most often determined in numerical simulations of lattice-regularised QCD.  A relevant comparison can be obtained with the results of Sect.\,\ref{mompade} by analysing the Fourier transform of Eq.\,(\ref{eq:simple_model}):
\begin{equation}
C(\tau) = b \, \delta(\tau) + \sum_{i=0}^{M-1} \frac{a_{i}} {2m_{i}}\, e^{-m_{i}\tau}\,, \label{eq:ft_simple_model}
\end{equation}
where $\tau$ is a Euclidean time variable.  Equation~(\ref{eq:ft_simple_model}) exhibits the behaviour expected of a Schwinger function in a channel with only stable bound states.

\subsubsection{Effective mass}
\label{sec:effective-mass}
For any observable projection of a two- or three-point Schwinger function the evolution at large Euclidean time is determined by the lowest mass state that couples to the channel under consideration.  One can therefore define an effective mass whose large-$\tau$ behaviour provides an approximation to the ground state mass:
\begin{equation}
m_{\rm eff}(\tau) := - \frac{1}{\delta\tau} \ln\left[\frac{C(\tau+\delta\tau)}{C(\tau)}\right] = m_0 + \mathcal{O}\left(e^{-(m_{1}-m_{0})\tau}\right)\,.
\label{eq:effective_mass}
\end{equation}

\begin{figure}[t]
\begin{center}
\centerline{\includegraphics[%
  clip,
  width=1.0\textwidth]{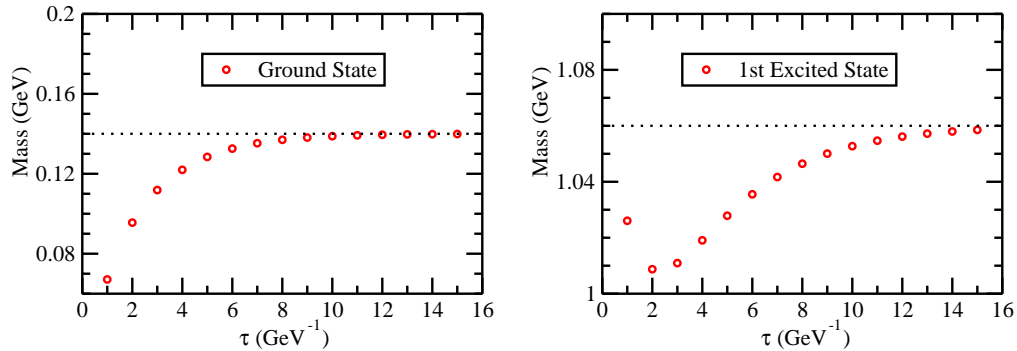}}

\caption{\label{fig:eff_mass}  Effective mass plots obtained from Eq.\,(\protect\ref{eq:effective_mass}).  \textit{Left panel} -- Calculated from Eq.\,(\ref{eq:ft_simple_model}) as written.  \textit{Right panel} -- Effective mass calculated with $a_0=0$ in Eq.\,(\protect\ref{eq:ft_simple_model}); i.e., with the lowest mass state omitted.  NB.\ $9 \, (m_1-m_0) \approx 15\, (m_2-m_1)$, Table \protect\ref{tab:Model_Param} -- cf.\ value of $\tau$ coinciding with the appearance of a plateau in each panel.  In both panels the horizontal dotted line indicates the relevant mass value, $m_0$ or $m_1$.}
\end{center}
\end{figure}

The effective mass for the Schwinger function in Eq.\,(\ref{eq:ft_simple_model}) is depicted in Fig.\,\ref{fig:eff_mass}.  It was obtained from $C(\tau)$ calculated at intervals $\delta \tau = 1\,{\rm GeV}^{-1}$ with $\tau_{\rm max}=16\,$GeV$^{-1} (>1/m_1)$.  This is a mockup of the situation in which $C(\tau)$ is only available on a discrete grid.  The grid spacing used here does not affect the illustration in this subsection.  As apparent in the figure, the rate at which $m_{\rm eff}(\tau)$ approaches the ground state's mass depends on $\Delta m = m_1-m_0$.\footnote{The effective mass approaches a plateau from below because $a_{i+1}/a_i < 0$ in Table~\ref{tab:Model_Param}.}  In principle, from a given Schwinger function the mass of the first excited state can be inferred via the $m_{\rm eff}(\tau)$-method if the ground state can be projected out.  To be successful this requires that: the ground state residue and mass be accurately determined; and the level separation between the first and second excited states is not significantly smaller than that between the ground and first excited state (a smaller separation requires that a larger domain of $\tau$ have a good signal-to-noise ratio).

\subsubsection{Exponential fits}
\label{sec:exponential_fits}
The last observation completed logically suggests that one might recover the bound state spectrum contained in a given Schwinger function by fitting a sum of exponentials \cite{mcneile}:  
\begin{equation}
\label{Esum}
E_{N_e}(\tau) = \sum_{k=1}^{N_e} \alpha_k \, {\rm e}^{- \mu_k \tau}.
\end{equation}
Table \ref{table:exp_fits} contains the results of a least-squares fitting procedure when applied to $C(\tau)$ in Eq.\,(\ref{eq:ft_simple_model}), sampled as described above.

This procedure is as efficient but not more so than that of Sect.\,\ref{mompade}.  The lowest mass state is reliably obtained and, as $N_e$ is increased, so is the next least massive state.  However, the remaining masses are false, even though the $R^2$ error is small.  This fact is not changed by choosing instead an absolute-error measure for $R^2$.  This is disconcerting but straightforward to understand: for nonzero error tolerance a least-squares nonlinear minimisation problem does not generally possess a unique solution.  Table~\ref{table:exp_fits} makes this plain empirically.

The results in Table~\ref{table:exp_fits} can be viewed as an attempt to extract the masses within the framework of Bayesian statistics.  The sample is fitted with a single exponential.  The result is then used as an input constraint on a fit with two exponentials and so on, logically, until $N_e=5$.  It is apparent that even with a subjective but accurate assumption about the masses of the two lightest states, nothing is gained in reliability of the estimates for the higher mass exponentials.

\begin{table}[t]
\caption{\label{table:exp_fits} Masses ($\mu_i$, in GeV) extracted via a least-squares fit of a sum of $N_e$ exponentials, Eq.\,(\ref{Esum}), to $C(\tau)$ in Eq.\,(\protect\ref{eq:ft_simple_model}) sampled at $N=16$ intervals of $\delta \tau = 1\,$GeV$^{-1}$ with $\tau_{\rm max}=16\,$GeV$^{-1}$.  $R^2= \sum_{i=1}^N [E_{N_e}(\tau_i)/C(\tau_i)-1]^2$.  The last row repeats, for convenience of comparison, the generating values for $C(\tau)$ (Table \protect\ref{tab:Model_Param}).}
\begin{center}
\begin{tabular*}{0.6\textwidth}{
|c@{\extracolsep{0ptplus1fil}}
|c@{\extracolsep{0ptplus1fil}}|c@{\extracolsep{0ptplus1fil}}|c@{\extracolsep{0ptplus1fil}}
|c@{\extracolsep{0ptplus1fil}}|c@{\extracolsep{0ptplus1fil}}|c@{\extracolsep{0ptplus1fil}}|}\hline
$N_e$  & 
$\mu_{0}$&
$\mu_{1}$&
$\mu_{2}$&
$\mu_{3}$&
$\mu_{4}$&
$\rule{0ex}{3ex} R^{2}$\tabularnewline
\hline
\hline 
2 &
0.14&
0.89&
---&
---&
---&
$\rule{0ex}{3ex}<10^{-06}$\tabularnewline
3 &
0.14&
1.05&
1.80&
---&
---&
$<10^{-10}$\tabularnewline
4 &
0.14&
1.06&
1.62&
1.97&
---&
$<10^{-11}$\tabularnewline
5 &
0.14&
1.06&
1.36&
1.73&
1.94&
$<10^{-11}$\tabularnewline
\hline 
5 &
0.14&
1.06&
1.72&
2.05&
2.2&
Source\tabularnewline
\hline
\end{tabular*}
\end{center}
\end{table}

\subsubsection{Error modelling}
\label{errormodel}
The analysis described hitherto is based on a curve that is sampled discretely but with near perfect accuracy.  This is the nature of DSE and kindred studies, in which the error is primarily systematic; i.e., it arises from truncation and, e.g., from uncertainties inherent in modelling the long-range part of the interaction between light-quarks.  However, in principle, once truncation and model are chosen, machine precision is achievable.  That is not so, e.g., with lattice simulations.  In addition to systematic errors, such as those associated with finite size and volume, quenching, extrapolation to realistic current-quark masses, etc., there is a material statistical error.  It is therefore interesting to generalise the model Schwinger function, Eq.\,(\ref{eq:ft_simple_model}), by adding Gaussian noise in the absolute value.

To be specific, we generate $M=100$ Gaussian-distributed random numbers $\{\mathcal{G}_J;J=1,\ldots,M\}$ with a mean of zero and a standard deviation of unity. Then from the exact values of $C(\tau)$ sampled at $N=16$ equally spaced points within $[0,\tau_{\rm max}]$: 
\begin{equation}
\{C_i=C(\tau_i)\,| \,\tau_i = i a, i=1,\ldots,N;\, a=\tau_{\rm max}/N\},
\end{equation}
we construct the sets $C^\varsigma_J$ with elements 
\begin{equation}
\label{varsigma}
(C^\varsigma_J)_i = C_i + \varsigma\,\mathcal{G}_J\, ; \; i = 1,\ldots,N\,.
\end{equation}
NB.\ Our reference value for the magnitude of the error is
\begin{equation}
\label{varsigma0}
\varsigma_0 = 0.01\,,
\end{equation}
which is $\approx C(2 N a)/C(0)$; i.e., the strength of the signal in the far infrared.

We thus obtain $M$ sets that are represented by the rows in the following array:
\begin{equation}
\label{M100sets}
\left\{
\begin{array}{cccc}
(C^\varsigma_1)_1 ,& (C^\varsigma_1)_2,& \ldots, &(C^\varsigma_1)_N\\
 (C^\varsigma_2)_1 ,& (C^\varsigma_2)_2,& \ldots,& (C^\varsigma_2)_N\\
\vdots & & & \\
 (C^\varsigma_M)_1 , & (C^\varsigma_M)_2,& \ldots, &(C^\varsigma_M)_N
\end{array}
\right\}
\end{equation}
Each column in this array, labelled by $i=1,\ldots,N$, may also be considered as a set, each of which we will call a distinct \textit{configuration}.  The configuration characterised by a given value of $i$ is an $M$-element set of \textit{measurements} of $C(\tau_i)$ with Gaussian standard deviation $\varsigma$ around a mean value $C_i$.

Reconsider now the effective mass defined in Eq.\,(\ref{eq:effective_mass}).  From each row of Eq.\,(\ref{M100sets}), define 
\begin{equation}
m_{J}^i = - \ln\left[\frac{(C^\varsigma_J)_{i+1}}{(C^\varsigma_J)_i}\right]\,,\; i=1,\ldots,N-1\,.
\end{equation}
This provides $M$ measurements of the effective mass for each time step; viz., the columns in 
\begin{equation}
\label{M100masses}
\left\{
\begin{array}{cccc}
m_1^1 ,& m_1^2,& \ldots, & m^{N-1}_1\\
m_2^1 ,& m_2^2,& \ldots,& m^{N-1}_2\\
\vdots & & & \\
m_M^1 , & m_M^2,& \ldots, &m_{M}^{N-1}
\end{array}
\right\}
\end{equation}
The average value of the effective mass at a given time step is  
\begin{equation}
\label{meanfit}
\bar m^i = \frac{1}{M} \sum_{J=1}^M m_J^i \,.
\end{equation}

The statistical error in $\bar m^i$ is determined via a jack-knife procedure, which yields a modification of the standard-deviation.  It is designed to account for correlations in the statistical fluctuations between consecutive measurements of a given quantity.  Here the quantity is $\bar m^i$ and we want to account for the possibility that if, e.g., $\bar m^i$ is above the true average then there is also a bias for $\bar m^{i+1}$ to be above the average.  If there are no such correlations in the errors then the method reproduces the standard deviation.

The procedure is implemented as follows.  For a given value of $i$ one considers the sample $\{m_J^i,J=1,\ldots,M\}$.  The element $m_1^i$ is omitted and one defines
\begin{equation}
\bar m_{\not\, 1}^i = \frac{1}{M-1} \sum_{J=2}^M m_J^i \,, \, i=1,\ldots,N-1\,.
\end{equation}
This is repeated, but with the $J=2$ element omitted instead, so that $m_{\not\, 2}^i$ is thereby obtained.  By stepping in this way through all $M$ elements one arrives at a set $\{ \bar m_{\not\, J}^i ; J = 1,\ldots{},M \} \,.$  The jack-knife estimate of the statistical error in $\bar m^i$ is then
\begin{equation}
\sigma^2_{\bar m^i} = \frac{M-1}{M} 
   \sum_{J=1}^M [\bar m_{\not\,J}^i-\bar m^i]^2 \,.
\end{equation}

We arrive in this way at a model for a data set that could arise via sampling of a system whose evolution is described by an effective mass $m_{\rm eff}(\tau)$; i.e., 
\begin{equation}
m_{\rm eff} = \{ \bar m_i \pm \sigma_{\bar m^i},\; i=1, \ldots{},N-1 \}\,.
\end{equation}
NB.\ While it is not immediately relevant, in the following we omit entries for which the error so calculated is greater than one-half of the mean value.

Our approach to mocking up a Gaussian statistical error is well defined.  It may nevertheless \emph{underestimate} the full extent of measurement errors.  For example, in a system with light-quarks one has for mesons \cite{mcneile} 
\begin{equation}
\frac{\rm signal}{\rm noise} \sim {\rm e}^{-(m_i - m_\pi)\tau},
\end{equation}
where $m_\pi$ is the pion mass, and this ratio is plainly small for meson masses not too much greater than $m_\pi$.  Moreover, one might find that in some channels there is a constant background noise, such as is the case in numerical simulations of glueball correlators \cite{richards}.   This possible defect in our method will, however, only tend to make us underestimate the impact of errors on the reliability of an extraction of information from a data sample.

\subsubsection{Results from noisy data}

\begin{figure}[t]
\begin{center}
\centerline{\includegraphics[%
  clip,
  width=1.0\textwidth]{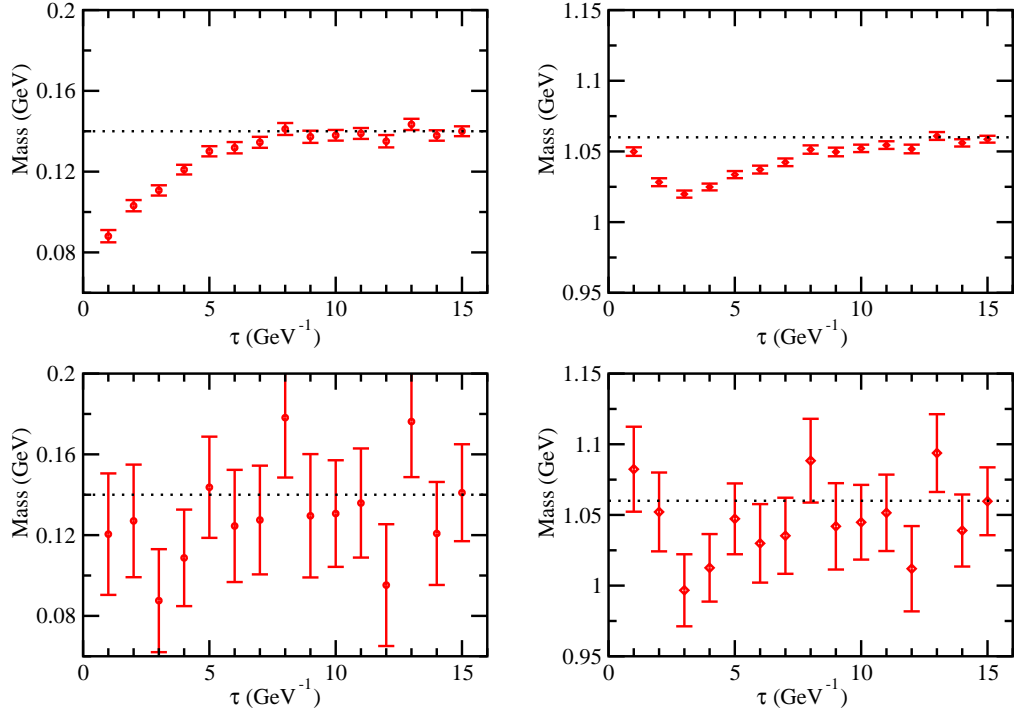}}
\caption{\label{fig:eff_mass_errors} Effective mass plots for a system characterised by $C(\tau)$ in Eq.\,(\protect\ref{eq:ft_simple_model}) with Gaussian noise added (Sect.\,\protect\ref{errormodel}, $M=100$ configurations): \textit{left panels}, ground state; and \textit{right panels}, first excited state, obtained with $a_0=0$ in Eq.\,(\protect\ref{eq:ft_simple_model}); i.e., with the lowest mass state omitted.  \textit{Upper panels} -- $\varsigma = 0.01\,\varsigma_0$, \textit{lower panels} -- $\varsigma = 0.1 \,\varsigma_0$.  In each panel the horizontal dotted line indicates the actual mass for the relevant state.}
\end{center}
\end{figure}

In Fig.\,\ref{fig:eff_mass_errors} we illustrate the ``noisy'' effective mass generated from $C(\tau)$ by the method just described.  This figure may be compared with Fig.\,\ref{fig:eff_mass} and the impact of the error is readily apparent.\footnote{It is noteworthy that our error model produces data which has the appearance of that obtained in numerical simulations of lattice-QCD (see, e.g., Refs.\,\protect\cite{Basak:2004hr,Morningstar:2005pv,Juge:2006gr,Graz}).  The magnitude of that error lies between our $\varsigma = 0.01\,\varsigma_0$ and $\varsigma = 0.1\,\varsigma_0$ samples.}  With $\varsigma$ just 10\% of $\varsigma_0$ a mass plateau is quite effectively obscured.  We subsequently work most often with $\varsigma = 0.01\varsigma_0$.  In this case we infer an effective mass by supposing that the domain $[\tau_9,\tau_{15}]$ lies within a mass plateau.  This yields mass values that are too low by only 1\%.  

\begin{figure}[t]
\begin{center}
\centerline{\includegraphics[%
  clip,
  width=1.0\textwidth]{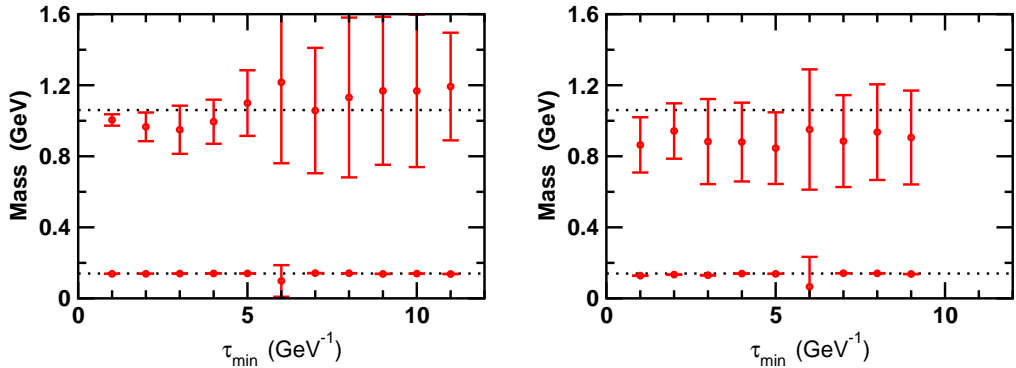}}
\end{center}
\vspace*{-6ex}

\caption{\label{fig:exp_sum_errors} $\tau_{\rm min}$-dependence in a least-squares fit of Eq.\,(\protect\ref{Esum}) to $C(\tau)$ in Eq.\,(\protect\ref{eq:ft_simple_model}) specified on $\tau\in [\tau_{\rm min},\tau_{16}]$ with Gaussian noise added (Sect.\,\protect\ref{errormodel}: $M=100$ configurations; $\varsigma=0.01\,\varsigma_0$; $R^2\lsim 10^{-5}$, uniformly).  \textit{Left panel} -- $N_e=3$; \textit{right panel} -- $N_e=4$.  Plateaux are visible in both cases.  In both panels the horizontal dotted lines indicate the relevant mass values, $m_0$ or $m_1$.}
\end{figure}

We also reconsider the method of Sect.\,\ref{sec:exponential_fits} in connection with the noisy sample.  In this case one performs an $N_e$ exponential least-squares fit to each one of the $M$ configurations in Eq.\,(\ref{M100sets}) and thereby determines the masses (and residues) for that configuration: $\{\mu^J_{k}$, $k=0,\ldots,(N_e-1)\}$.  The mean fit-masses, $\bar \mu_{k}$, are evaluated via Eq.\,(\ref{meanfit}) and the jack-knife procedure employed thereafter to determine the statistical error in each of these.  Results obtained for $\bar \mu_0,\bar\mu_1$ via $N_e=3,4$ exponential fits are depicted in Fig.\,\ref{fig:exp_sum_errors}.  Within errors, they are consistent with those reported in Table~\ref{table:exp_fits}.  This figure contains additional information; namely, it illustrates the dependence of the extracted mass on the domain of $\tau$ for which information is available.  We performed an exponential fit on the domain $\tau\in [\tau_{\rm min},\tau_{\rm max}=\tau_{16}]$, with $\tau_{\rm min}=\tau_1$, and then repeatedly incremented $\tau_{\rm min}$ until the number of fit parameters exceeded the information sample.  This occurs at $\tau_{\rm min}= \tau_{11}$ for $N_e=3$ (six pieces of information to fit six parameters) and $\tau_{\rm min}=\tau_9 $ for $N_e = 4$.  The value of $\tau_{\rm min}$ does not strongly influence the extracted value of $\bar \mu_0$.  However, the error in $\bar \mu_1$ tends to grow with $\tau_{\rm min}$.  This is not surprising because the ground state comes increasingly to dominate the signal at larger values of $\tau$.

These two examples suggest that the analysis of data with a Gaussian error of $\varsigma= 0.01\,\varsigma_0$ will yield results whose reliability is not significantly poorer than that obtained from a noiseless sample.  Naturally, an improved result cannot be obtained from noisier data.  Nevertheless, the techniques we have hitherto illustrated are patently inadequate to the task of inferring reliably anything but the lightest two masses.  A more advanced tool is necessary.

\section{Masses from Schwinger Functions: Correlator Matrix Method}
\label{correlatormatrixmethod}
A more sophisticated method follows from an appreciation that the simultaneous analysis of Schwinger functions obtained from numerous carefully chosen interpolating fields can assist in determining the properties of states other than the one with the lowest mass in a given channel \cite{Michael:1985ne,Luscher:1990ck}.  This procedure will naturally become very time consuming as the number of operators is increased.  

To introduce the method \cite{Leinweber:2004it,Juge:2006gr,Graz}, suppose one has operators $\{\chi_i(\vec{x},\tau),i=1,\ldots,K\}$ that each have a nonzero residue in a particular channel and for which one has available the correlation matrix
\begin{equation}
G_{ij}(\tau,\vec{p}) = \int d^3x\, {\rm e}^{-i \vec{p}\cdot \vec{x}}\langle0|\chi_{i}(\vec{x},\tau) \overline{\chi}_{j}(\vec{0},0)|0\rangle\,.
\end{equation}
Now insert a supposedly complete set of colour-singlet states whose quantum numbers characterise the channel of interest
\begin{equation}
\sum_{n=0}^{N_{\rm states}-1} \, | H_n \rangle \langle H_n| = \mathbf{1}\,,
\end{equation}
where momentum and other labels are not written explicitly, and it may be that $N_{\rm states}=\infty$.  One then has, using translational invariance of the vacuum,
\begin{equation}
\label{eq:variational_G}
G_{ij}(\tau) := G_{ij}(\tau,\vec{p}=0) = \sum_{n=0}^{N_{\rm states}-1} \, {\rm e}^{- m_n \tau}\, \ell^n_i \bar\ell^n_j
\end{equation}
with
\begin{equation}
\ell^n_i= \langle0| \chi_{i}(\vec{0},0) | H_n\rangle \,,\;  \bar\ell^n_j= \langle H_n | \bar\chi_{j}(\vec{0},0)| 0 \rangle\,.
\end{equation}
Equation\,(\ref{eq:variational_G}) is a many-operator analogue of Eq.\,(\ref{eq:ft_simple_model}) in Sect.\,\ref{sec:coord_space}.  NB.\ In the real-world application of this method, such as in numerical simulations of lattice-QCD, considerable effort must be devoted to the construction of the operators $\chi_i$ in order to optimise for success.  The operators are typically combinations of nonlocal few-quark/antiquark operators built and tuned empirically so as to have large overlap with states of interest in a given channel \cite{Leinweber:2004it,Juge:2006gr,Graz}.  Our analysis will highlight the importance of this step.

The correlation matrix $G(\tau)$ is Hermitian and hence one can solve the eigenvalue problem
\begin{equation}
\label{PrincipalC}
\left[G(\tau_0)^{-1/2} \, G(\tau) \, G(\tau_0)^{-1/2}\right] \phi^k(\tau,\tau_0) = \lambda^k(\tau,\tau_0)\, \phi^k(\tau,\tau_0) 
\end{equation}
to obtain $K$ ``principal correlators'' $\lambda^k(\tau,\tau_0)$.\footnote{$\tau_0$ is a reference time that is typically assigned a small value.  Reliable results should be independent of $\tau_0$.  We use $\tau_0=0$.}  With the operators $\chi_i$ defined and ordered such that $\lambda^{k} > \lambda^{k+1}$, it can then be shown that 
\begin{equation}
\lim_{\tau\to\infty} \lambda^k(\tau,\tau_0) = {\rm e}^{-(\tau - \tau_0)m^k} \left( 1 + {\rm O}({\rm e}^{-\tau \Delta m^k})\right)\,,\; \Delta m^k = {\rm min}_{j\neq k}|m^j - m^k|\,.
\end{equation}
One therefore has $K$ principal effective masses
\begin{equation}
m_{\rm eff}^{l}(\tau,\tau_0) = 
- \ln\left[ \frac{\lambda^{l+1}(\tau+\delta\tau,\tau_0)}{\lambda^{l+1}(\tau,\tau_0)}\right] \,,\;l=0,\ldots,(K-1),
\end{equation}
each of which will, for large $\tau$, exhibit a single plateau that corresponds to the mass of one of the $K$ lowest-mass states in the channel under consideration.  In practice $K \ll N_{\rm states}$.

We now explore and illustrate the efficacy of this method.  Suppose there is a colour-singlet channel that possesses a trajectory of bound-states whose masses are given by \cite{Anisovich:2000kx}:
\begin{equation}
m_{l}^{2} = m_{0}^{2} + l \mu^{2}\,, l=0,1,\ldots,(l_{\rm max}-1) \,, \label{eq:regge_trajectory}
\end{equation}
where $a_\tau m_{0}=0.5$, $a_\tau^2 \mu^{2}=0.36$ and the length-scale $a_\tau=0.1\,$fm.  We will subsequently work with $l_{\rm max} = 20$.  

\subsection{Perfect operators}
As a first step we suppose that a correlator matrix of the form in Eq.\,(\ref{eq:variational_G}) has been obtained with $K=6$ operators, and assume in addition that these operators are \emph{perfect}; namely, each operator produces a unique state from the vacuum, so that 
\begin{equation}
\label{perfectchi}
\ell_{i}^{n}= \delta_{i(n+1)} \,.
\end{equation}
The correlator matrix is therefore
\begin{equation}
\label{Gperfect}
G(\tau) = 
\left( \begin{array}{cccccc}
{\rm e}^{- m_0 \tau} & 0 & 0 & 0 & 0 & 0 \\
0 & {\rm e}^{- m_1 \tau} & 0 & 0 & 0 & 0 \\
0 & 0 & {\rm e}^{- m_2 \tau} &  0 & 0 & 0 \\
0 & 0 & 0 & {\rm e}^{- m_3 \tau} & 0 & 0 \\
0 & 0 & 0 & 0 & {\rm e}^{- m_4 \tau} & 0 \\
0 & 0 & 0 & 0 & 0 & {\rm e}^{- m_5 \tau}\\
\end{array}
\right).
\end{equation}
This situation cannot be achieved in reality because it requires \textit{a priori} complete information about all bound states in a given channel and presumes that all the states are stable.  It is nevertheless a useful demonstration case.  

To this correlator matrix we add Gaussian error in the manner described in Sect.\,\ref{errormodel}; namely, the matrix is treated in the same way as the function in Eq.\,(\ref{eq:ft_simple_model}).  A given element is sampled at $N=16$ time steps, measured here in units of $a_\tau$, to produce $\{ G_{ij}(s a_\tau), s=1,\ldots,N\}$.  We then generate $M=100$ Gaussian-distributed random numbers $\{\mathcal{G}_J;J=1,\ldots,M\}$ with a mean of zero and a standard deviation of unity, and thereafter construct the sets $\{G_{ij}(s)\}^\varsigma_J$ with elements 
\begin{equation}
\label{varsigmaG}
[G_{ij}(s)]^\varsigma_J = G_{ij}(s) + \varsigma\, \mathcal{G}_J\, ; \; s = 1,\ldots,N\,.
\end{equation}
This yields $M$ correlator matrices sampled at $N$ $\tau$-values, which appear as the rows in the following matrix:
\begin{equation}
\label{M100Gij}
\left\{
\begin{array}{cccc}
G(1)^\varsigma_1, & G(2)^\varsigma_1,& \ldots, & G(N)^\varsigma_1,\\
G(1)^\varsigma_2, & G(2)^\varsigma_2,& \ldots, & G(N)^\varsigma_2,\\
\vdots & & & \\
G(1)^\varsigma_M, & G(2)^\varsigma_M,& \ldots, & G(N)^\varsigma_M,
\end{array}
\right\}
\end{equation}
(NB. Each entry is a $K\times K$ matrix.)  As before, each column in this array, labelled by $s=1,\ldots,N$, may also be considered as a set, each of which is a distinct \textit{configuration}.  The configuration identified by a given value of $s$ is an $M$-element set of \textit{measurements} of the correlator matrix $G(s)$ with Gaussian standard deviation $\varsigma$ around a mean value $G(s)$.

From each element in a given row of the matrix in Eq.\,(\ref{M100Gij}), the $K$ principal correlators can be determined by solving Eq.\,(\ref{PrincipalC}), and therefrom the principal effective masses:
\begin{equation}
\label{meffs}
m_{J}^{l}(s) = 
- \ln\left[ \frac{\lambda_J^{l+1}(s+1,s_0)}{\lambda_J^{l+1}(s,s_0)}\right] , \; l=0,\ldots,(K-1)\,.
\end{equation}
This provides $M$ measurements of each of the principal effective masses at each time step $\tau (= s a_\tau)$; i.e., the columns in 
\begin{equation}
\label{M100Pmasses}
\left\{
\begin{array}{cccc}
m_1^l(1) ,& m_1^l(2),& \ldots, & m_1^l(s-1)\\
m_2^l(1) ,& m_2^l(2),& \ldots, & m_2^l(s-1)\\
\vdots & & & \\
m_M^l(1) ,& m_M^l(2),& \ldots, & m_M^l(s-1)
\end{array}
\right\}\,, \; l = 0,\ldots,(K-1),
\end{equation}
and hence an average value of a principal effective mass at a given time step:
\begin{equation}
\label{meanfitk}
\bar m^l(s) = \frac{1}{M} \sum_{J=1}^M m_J^l(s) \,.
\end{equation}
The statistical error in $\bar m^l(s)$ is calculated via the jack-knife procedure described in Sect.\,\ref{errormodel}

\begin{figure}[t]
\centerline{\includegraphics[%
  clip,
  width=1.0\textwidth]{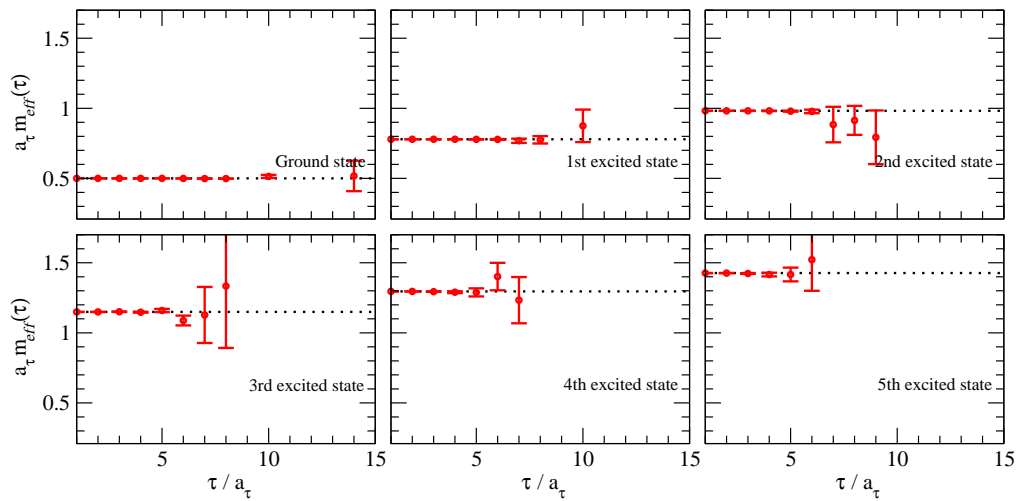}}

\caption{\label{fig:perfect_cor_mat} Effective mass plots from analysis of the correlator matrix in Eq.\,(\protect\ref{Gperfect}) with Gaussian noise added via $M=100$ configurations with $\varsigma=0.01\,\varsigma_0$.  In each panel the horizontal dotted-line indicates the input mass, Eq.\,(\protect\ref{eq:regge_trajectory}).}
\end{figure}

The results of this analysis are presented in Fig.\,\ref{fig:perfect_cor_mat}.  With perfect operators a plateau at the correct mass [input, Eq.\,(\protect\ref{eq:regge_trajectory})] is immediately apparent in each panel, a feature that illustrates the power of the method.  The signal is degraded through noise as $\tau$ increases but the impact of that is minimal when the plateaux are so readily evident.  

It will be observed that in each panel some time steps are omitted and, for all but the ground state, effective mass values are not plotted for $\tau$ greater than some minimum value.  The omission occurs because Gaussian noise can lead to $\lambda(s+1)/\lambda(s)>1$ in which case Eq.\,(\ref{meffs}) cannot define a positive mass.  This fact and our decision to omit data for which the error is greater than one-half of the mean value explains the absence of points at larger $\tau$.\footnote{Such features are also apparent in the data obtained from numerical simulations of lattice-QCD, e.g., Refs.\,\protect\cite{Juge:2006gr,Graz}.}

\subsection{Untuned cf.\ optimised operators}
It is impossible in principle to begin with perfect operators, $\chi_i$.  Thus to simulate a more realistic situation involving the determination of a three-point Schwinger function we write
\begin{equation}
\label{linR}
\ell_i^n = (-1)^{n} \, | \mbox{r}_i |\,,\; i=1,\ldots,K\,,
\end{equation}
where ``r$_i$'' is a number chosen randomly from a normal distribution with mean zero and variance one.\footnote{As noted in Ref.\,\protect\cite{Holl:2004fr}, the coefficients of states in a physical three-point Schwinger function must alternate in sign.  This is not the case for a four-point polarisation function.}  We subsequently normalise such that
\begin{equation}
\label{linRN}
\sum_{n=0}^{N_{\rm states}-1} |\ell_i^n|^2 = 1\,, \; i=1,\ldots,K\,,
\end{equation}
where in our concrete examples $N_{\rm states}=l_{\rm max}$.  Normalisation does not compromise generality and is a useful precaution against precocious noise.  Equations (\ref{linR}), (\ref{linRN}) specify a rank $K=6$ correlator matrix associated with a channel in some system that contains $l_{\rm max}=20$ bound states.  To this matrix we added Gaussian noise, as in the case of perfect operators, and repeated the analysis that led to Fig.\,\ref{fig:perfect_cor_mat}.  

\begin{figure}[t]
\begin{center}
\centerline{\includegraphics[%
  clip,
  width=1.0\textwidth]{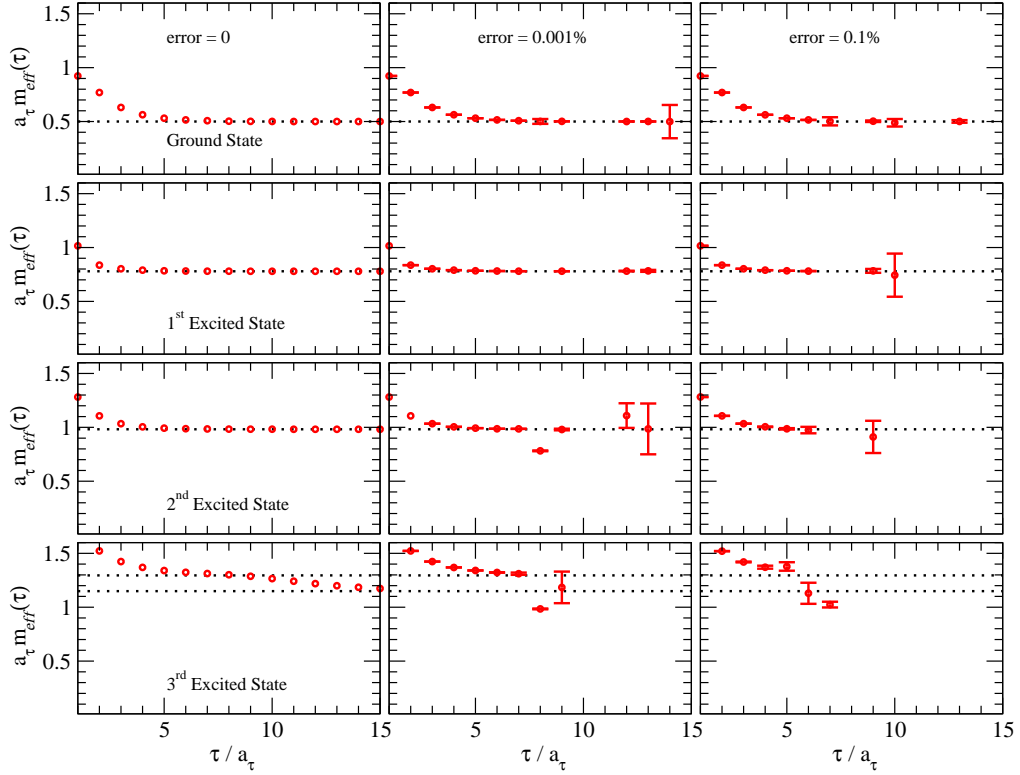}}
\end{center}

\caption{\label{fig:default_corr_mat} Principal effective masses from analysis of the untuned correlator matrix described in connection with Eqs.\,(\protect\ref{linR}), (\protect\ref{linRN}).  In this case the operators $\{\chi_i,i=1,\ldots,K\}$ are not perfect.  \textit{Column 1} -- Precise measurement of the correlator matrix; i.e., Gaussian noise $\varsigma = 0$; \textit{Column 2} -- Gaussian error of $\varsigma = 10^{-5}\,\varsigma_0$; and  \textit{Column 3} -- Gaussian error of $\varsigma = 10^{-3}\,\varsigma_0$.  The horizontal dotted-line in the top nine panels indicates the relevant input mass; viz., $m_{0,1,2}$ from Eq.\,(\protect\ref{eq:regge_trajectory}).  In each of the bottom three panels there are two horizontal lines: the lower is $m_3$, which is the correct result, and the upper is $m_4$ in order to provide a context for any putative plateau.}
\end{figure}

The results are presented in Fig.\,\ref{fig:default_corr_mat}.  It is evident from column-one that the method is far less reliable with untuned operators since even a perfect measurement of the correlator matrix is unable to provide an accurate result for the third excited state: there is no true plateau.  Furthermore, for this state the presence of even an extremely small amount of Gaussian noise in the correlator matrix eliminates the possibility of obtaining any information: the signal is lost altogether with a Gaussian error of only $\varsigma=0.1\% \, \varsigma_0=0.001\,\varsigma_0=0.00001$.

Our illustrations have shown that tuning the operators is an important step in the practical application of the correlator matrix method.  This operation may be represented by emphasising the diagonal elements in the correlator matrix but allowing for leakage into the off-diagonal elements.  To be specific, we write
\begin{equation}
\label{tunedG}
\ell^n_i  = \left\{
\begin{array}{cc}
\displaystyle \sqrt{y} \, , & i = n+1 \\ 
\rule{0em}{5ex}
\displaystyle \sqrt{\frac{1-y}{(N_{\rm states}-1)}} \, |\mbox{r}_i|\,, & i \neq n+1\,,
\end{array} 
\right.
\end{equation}
where again ``r$_i$'' is a number chosen randomly from a normal distribution with mean zero and variance one, and subsequently normalise.  A value of $y=0.5$ is described as 50\% optimisation.  This procedure specifies a correlator matrix to which we add Gaussian noise.  

\begin{figure}[t]
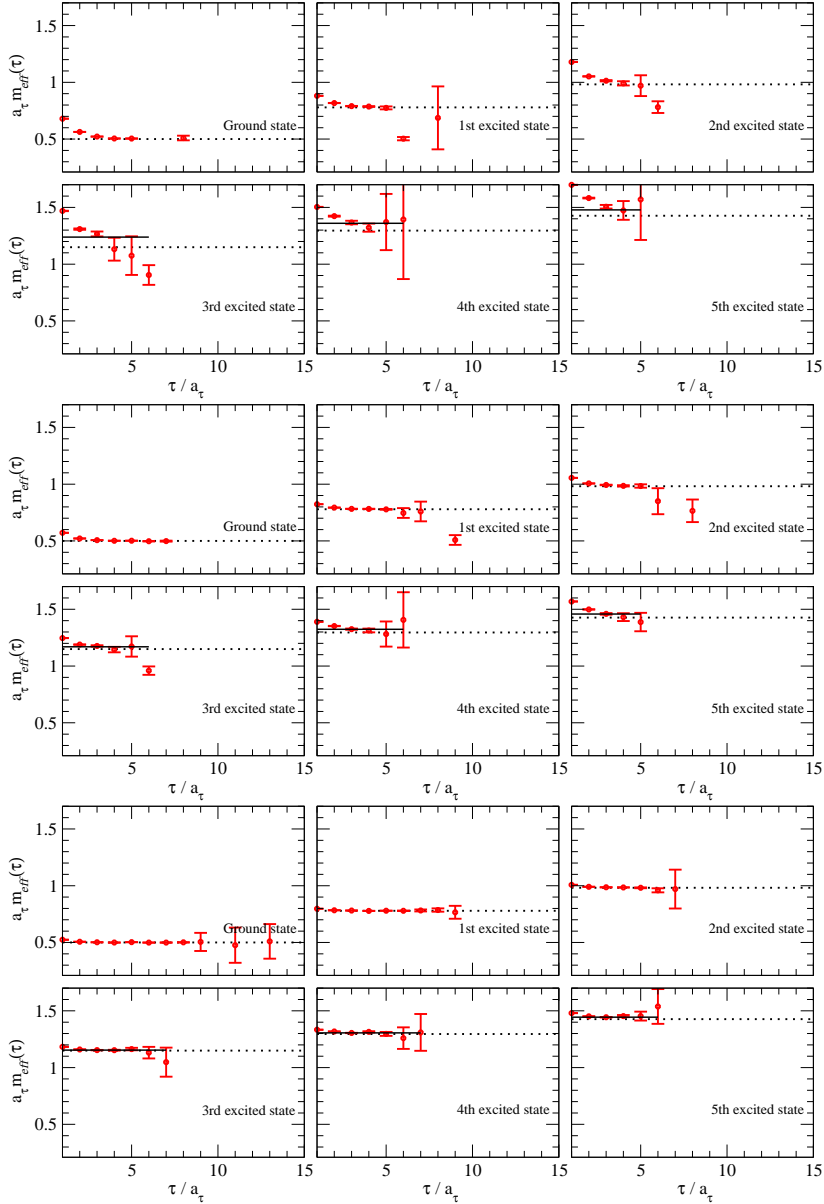

\begin{center}
\begin{minipage}[c]{0.81\textwidth}%
\includegraphics[%
  clip,
  width=1\textwidth]{Fig12u.eps}\\
\includegraphics[clip,width=1\textwidth]{Fig12m.eps}\\
\includegraphics[%
  clip,
  width=1\textwidth]{Fig12l.eps}
\end{minipage}
\vspace*{-3ex}

\end{center}
\caption{\label{fig:spread_1e_2} Principal effective masses for the six lightest states generated by Eq.\,(\protect\ref{eq:regge_trajectory}) calculated from the tuned-operator correlator matrix described in connection with Eq.\,(\protect\ref{tunedG}) (Gaussian noise: $M=100$ configurations; $\varsigma=0.01\,\varsigma_0$):
\textit{Upper 6 panels} -- 50\% optimisation;
\textit{Middle 6} -- 70\%;
\textit{Bottom 6} -- 90\%.  The horizontal dotted-line indicates the relevant input mass.  Where it appears, the solid line is an error-weighted least-squares fit to the data beginning with the point at $\tau =3 a_\tau$.}
\end{figure}

The results for the principal effective masses are depicted in Fig.\,\ref{fig:spread_1e_2}.  With 50\% optimisation a clear plateau is apparent for the lowest mass state and, arguably, for the next lightest state, but beyond that one cannot confidently locate a plateau.  Moreover, if one nevertheless chooses to estimate masses from defined rather than evident plateaux in the data, those masses are degenerate within errors.  However, for 70\% or better optimisation, clear plateaux are distinguishable for all of the six lightest states and accurate masses can be inferred. 

\begin{figure}[t]
\begin{center}\begin{minipage}[c]{1.0\textwidth}%
\includegraphics[%
  clip,
  width=1\textwidth]{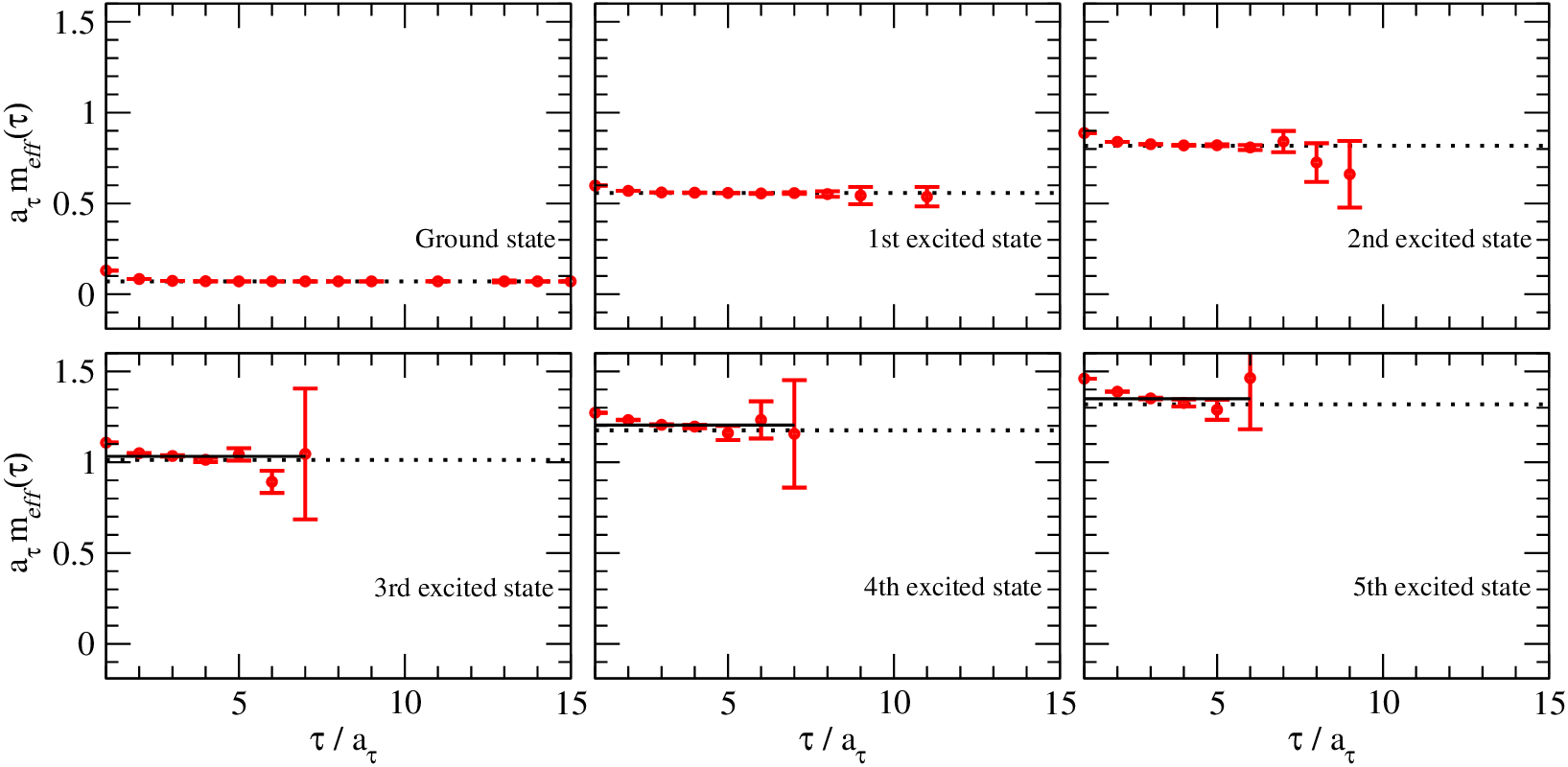}\\
\includegraphics[%
  clip,
  width=1\textwidth]{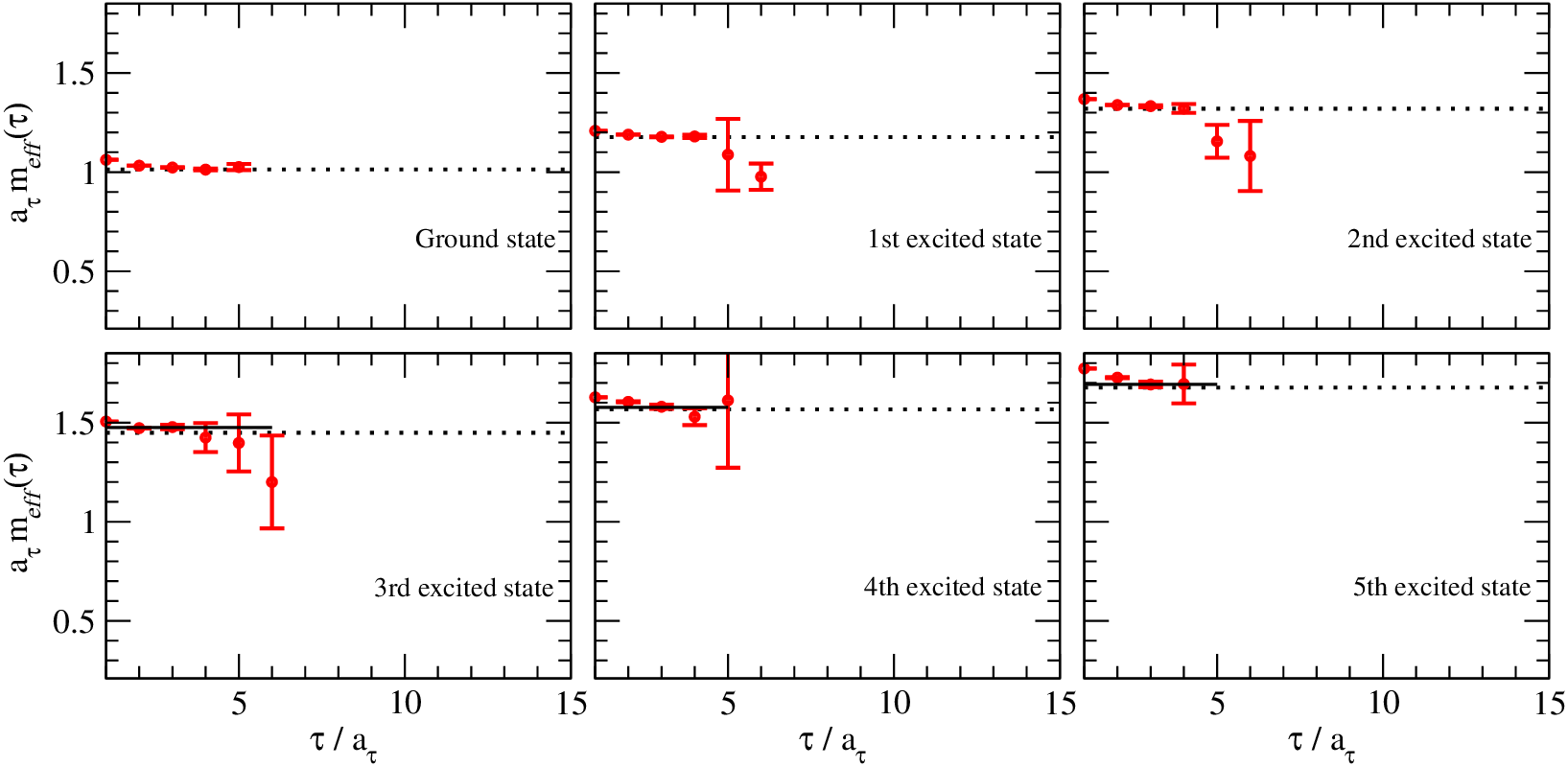}\end{minipage}%
\end{center}

\caption{\label{fig:m0_spread_1e_2} Principal effective masses obtained from a correlator matrix for the six lightest states generated by the formula in Eq.\,(\protect\ref{eq:regge_trajectory}): \textit{upper 6 panels} -- $a_\tau m_0 = 0.071$; \textit{lower 6 panels} -- $a_\tau m_0 = 1.02$.  In all panels the horizontal dotted-line indicates the relevant input mass.  Where it appears, the solid line is an error-weighted least-squares fit to the data beginning with the point at $\tau =3 a_\tau$.  (Tuned correlator matrix with 80\% overlap, Eq.\,(\protect\ref{tunedG}). Gaussian noise: $M=100$ configurations; $\varsigma=0.01\,\varsigma_0$).}
\end{figure}

\subsection{Influence of level location}
\begin{figure}[t]
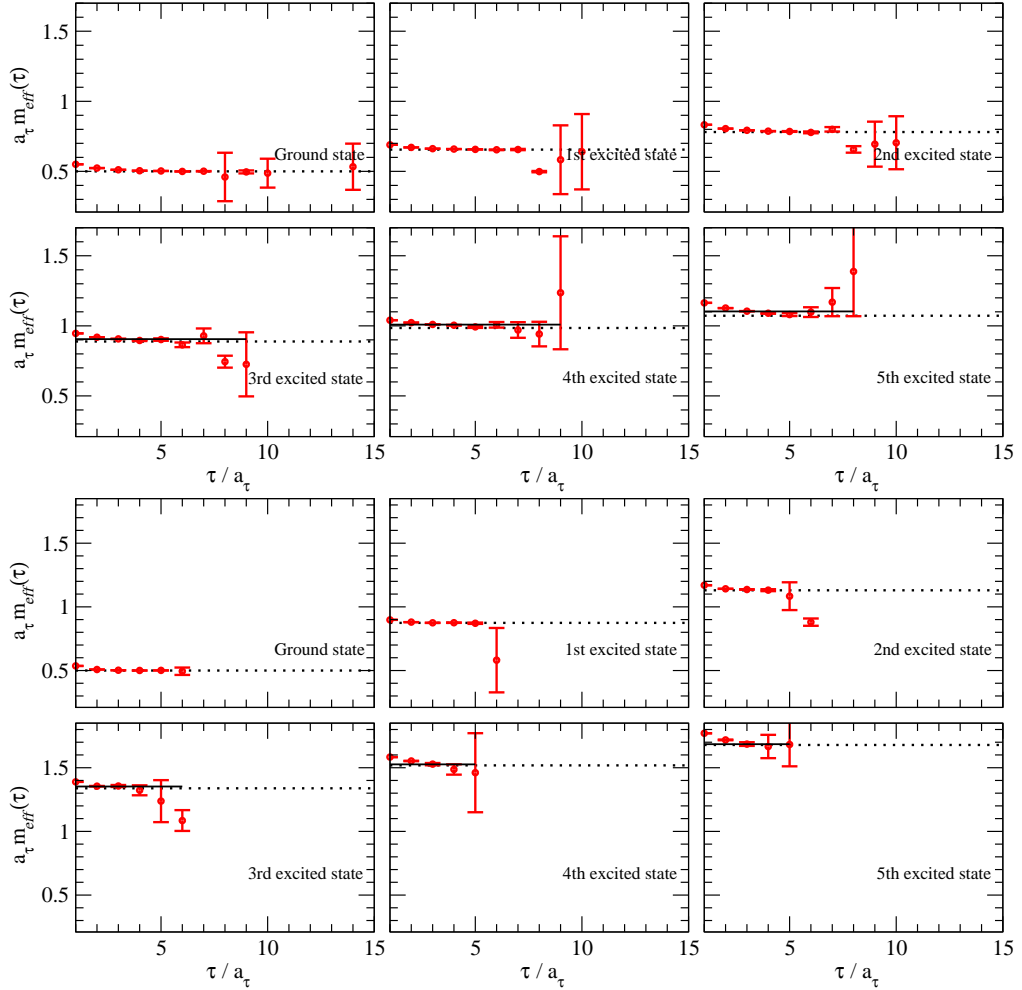

\begin{center}\begin{minipage}[c]{1.0\textwidth}%
\includegraphics[%
  clip,
  width=1\textwidth]{Fig14a.eps}\\
\includegraphics[%
  clip,
  width=1\textwidth]{Fig14b.eps}
\end{minipage}
\end{center}
\caption{\label{fig:slope_spread_1e_2} Principal effective masses obtained from a correlator matrix for the six lightest states generated by the formula in Eq.\,(\protect\ref{eq:regge_trajectory}): \textit{upper 6 panels} -- $a_\tau^2\mu^2 = 0.18$; \textit{lower 6 panels} -- $a_\tau^2\mu^2 = 1.02$.  In all panels the horizontal dotted-line indicates the relevant input mass.  Where it appears, the solid line is an error-weighted least-squares fit to the data beginning with the point at $\tau =3 a_\tau$.  (Tuned correlator matrix with 80\% overlap, Eq.\,(\protect\ref{tunedG}).  Gaussian noise: $M=100$ configurations; $\varsigma=0.01\,\varsigma_0$).}
\end{figure}

Two other features of the spectrum in a given channel may reasonably be expected to impact upon the extraction of masses from a Schwinger function; namely, the position of the ground state and the level separation.  One may investigate these effects by varying the parameters in Eq.\,(\ref{eq:regge_trajectory}).  To explore the first, in Fig.\,\ref{fig:m0_spread_1e_2} we compare $a_\tau m_0 = 0.071$ ($m_0 \simeq 0.14\,$GeV) with $a_\tau m_0 = 1.02$ ($m_0 \simeq 2.0\,$GeV), while in Fig.\,\ref{fig:slope_spread_1e_2} we contrast $a_\tau^2\mu^2 = 0.18$ with $a_\tau^2\mu^2 = 1.02$.  NB.\ For these comparisons we used a tuned correlator matrix constructed according to Eq.\,(\ref{tunedG}) with 80\% optimisation.

It is evident in Fig.\,\ref{fig:m0_spread_1e_2} that the primary effect of an increase in $a_\tau m_0$ is a reduction, for all states, in the domain on which a plateau may be observed.  The larger value of $a_\tau m_0$ causes the signal to more rapidly enter the domain on which noise becomes dominant.  Figure~\ref{fig:slope_spread_1e_2} illustrates that similar statements apply when changes in $a_\tau^2 \mu^2$ are considered.  It follows that for any given channel, ground state mass and level spacing, smaller values of $a_\tau$ will help improve the method's accuracy if $m_0 \tau_{\rm max}$, $\tau_{\rm max} = N a_\tau$, can be held fixed and noise levels kept constant.  Satisfying these requirements simultaneously is challenging.

\section{Epilogue}
\label{epilogue}
We analysed the capacity of Schwinger functions to yield information about bound states, and established that for the ground state in a given channel the mass and residue are accessible via rudimentary methods.  When the mass-splitting between the first and second excited states is not significantly smaller than that between the ground and first excited states, these methods may also reliably provide the same information about the first excited state, so long as the Schwinger function under consideration does actually possess the analytic properties assumed of it.  However, simple methods cannot provide dependable information about more massive states in a given channel.

Indeed, there is no easy way to extract such information.  An approach based on a correlator matrix can be successful but only if the operators are carefully constructed so as to have large overlap with states of interest in a given channel and statistical errors can be made small; viz., $\sim 1$\%.  While it is possible in principle to satisfy these constraints, doing so is labour intensive and time consuming.  That is nevertheless justified in the absence of model-dependence.

We have posed the question of whether, in the context of bound-state studies in which model assumptions are made regarding the nature of the long-range interaction between light quarks, labour can be saved and/or accuracy gained by working solely with Schwinger functions.  In our view the analysis herein suggests strongly that the answer is \emph{no}.  In their formulation the studies of this type explicitly define an analytic continuation of a model into the complex plane.  Hence all momenta are directly accessible.  We saw clearly that it is not possible to avoid model artefacts by pretending ignorance of this.  Hence nothing is gained by solving a complex of DSEs for the Schwinger functions associated with numerous carefully tuned interpolating fields.  Moreover, capitalising on the feature that all momenta are directly accessible is the most efficient manner by which to test and improve the model input, and thereby to relate experimental data on bound state properties to information about the interaction between light quarks.  This is further emphasised by the fact that because excited states are unstable they are characterised by complex pole positions.

\begin{acknowledge}
We acknowledge profitable interactions with P.\,O.~Bowman, R.\,G.~Edwards, C.~Gattringer, D.\,B.~Leinweber, P.~Maris, P.\,C.~Tandy and A.\,W.~Thomas.  SVW thanks the staff and students in the Special Research Centre for the Subatomic Structure of Matter (CSSM) at the University of Adelaide for their hospitality during a visit in which part of this work was performed. 
This work was supported by: Department of Energy, Office of Nuclear Physics, contract no.\ DE-AC02-06CH11357; 
the Austrian Science Fund FWF, Schr\"odinger-R\"uckkehrstipendium R50-N08; 
and benefited from the ANL Computing Resource Center's facilities.
\end{acknowledge}

\appendix
\section{Pole residue in a vertex} 
\setcounter{section}{1}
\label{App:residue} 
Equation~(\ref{eq:DSE_inhomogeneous}) is the inhomogeneous pseudoscalar Bethe-Salpeter equation (BSE).  The general solution has the form in Eq.\,(\ref{genpvv}).  Each scalar function therein depends on the renormalisation point but we have not written that explicitly.  The inhomogeneous equation is solved subject to a renormalisation boundary condition:
\begin{equation}
\label{renorm}
\left.\Gamma_{5}^j(k;P)\right|_{k^2=\zeta^2,P=0} = \frac{1}{2} i \gamma_5 \tau^j\,,
\end{equation}
which translates into the single condition
\begin{equation}
\left. E_{5}(k;P)\right|_{k^2=\zeta^2,P=0} = 1.
\end{equation}
This result is guaranteed so long as the renormalisation constant for the pseudoscalar vertex is the same as that for the Lagrangian mass; viz., 
\begin{equation}
\label{Z4Z5}
Z_5(\zeta^2,\Lambda^2) = Z_4(\zeta^2,\Lambda^2) .
\end{equation}
This is a consequence of the axial-vector Ward-Takahashi identity.

In the neighbourhood of any pseudoscalar meson pole,
\begin{equation}
\left. \Gamma_{5 }^j(k;P)\right|_{P^2+m_{\pi_n}^2 \approx 0}
=   \frac{\rho_{\pi_n} }{P^2 + 
m_{\pi_n}^2} \Gamma_{\pi_n}^j(k;P)\,, \label{genpv} 
\end{equation}
where 
\begin{equation}
\Gamma_{\pi_n}^j(k;P) = \tau^j \gamma_5 \left[ i E_{\pi_n}(k;P) + \gamma\cdot P F_{\pi_n}(k;P) +
    \gamma\cdot k \,k \cdot P\, G_{\pi_n}(k;P) + 
\sigma_{\mu\nu}\,k_\mu P_\nu \,H_{\pi_n}(k;P)  \right] \!  \label{genpibsa} 
\end{equation}
is the canonically normalised bound-state Bethe-Salpeter amplitude, and 
\begin{equation}
\label{rhon}
i  \rho_{\pi_n}\!(\zeta^2)\,   = Z_4(\zeta^2,\Lambda^2)\,{\rm tr} 
\int^\Lambda_q \gamma_5 \, \chi_{\pi_n}(q;P)
\end{equation}
is the gauge-invariant and regularisation-scale-independent pseudoscalar projection of the meson's Bethe-Salpeter wave function onto the origin in configuration space.  In Eq.\,(\ref{rhon}),  $\chi_{\pi_n}$ is specified analogously to Eq.\,(\ref{chidef}) but with the inhomogeneous vertex replaced by Eq.\,(\ref{genpibsa}).  Here, as elsewhere herein, the suffix $n$ labels nondegenerate states in the channel: $n=0$ is the ground state, $n=1$ is the first excited state, etc.

In the isospin symmetric limit, Eq.\,(\ref{genpv}) entails 
\begin{equation}
{\rm tr}_F \, \tau^i \Gamma_{5}^j(k;P) = \delta^{ij} \Gamma_5(k;P) = \delta^{ij}\frac{2 \rho_{\pi_n} }{P^2 + 
m_{\pi_n}^2} \Gamma_{\pi_n}(k;P)
\end{equation}
so that in the neighbourhood of a pole \begin{equation}
\left. \Gamma_5(0;P)\right|_{P^2+m_{\pi_n}^2 \approx 0}= \frac{2 \rho_{\pi_n} }{P^2 + 
m_{\pi_n}^2} \Gamma_{\pi_n}(0;P)\,
\end{equation}
or, equivalently, 
\begin{equation}
\label{residue}
E_5(0;P) = \frac{1}{P^2 + 
m_{\pi_n}^2} \, 2 \rho_{\pi_n} E_{\pi_n}(0;P)\,.
\end{equation}

The residue of the pole in the vertex is therefore
\begin{equation}
\label{residueform}
r_{E_{\pi_n}}= 2 \rho_{\pi_n} E_{\pi_n}(0;P)\,,
\end{equation}
which can be calculated so long as one has the solution of the homogeneous BSE in addition to the solution of Eq.\,(\ref{eq:DSE_inhomogeneous}).  Thus do we calculate the dotted curves in the right panels of Figs.\,\ref{fig:ps_meson_mass_residue} and \ref{fig:scalar}.



\begin{thebibliography}{99}

\bibitem{glimm} J.~Glimm and A.~Jaffee, \textit{Quantum Physics.\ A Functional Point of View} (Springer-Verlag, New York, 1981).

\bibitem{hawes} F.\,T.~Hawes, P.~Maris and C.\,D.~Roberts,
  Phys.\ Lett.\ B {\bf 440}, 353 (1998).

\bibitem{mandarquark} M.\,S.~Bhagwat, M.\,A.~Pichowsky, C.\,D.~Roberts and P.\,C.~Tandy,
  Phys.\ Rev.\ \textbf{C\,68}, 015203 (2003).
  
\bibitem{fischergluon} R.~Alkofer, C.\,S.~Fischer, H.~Reinhardt and L.~von Smekal,
  Phys.\ Rev.\ \textbf{D\,68}, 045003 (2003)

\bibitem{detmoldquark} R.~Alkofer, W.~Detmold, C.\,S.~Fischer and P.~Maris,
  Phys.\ Rev.\ \textbf{D\,70}, 014014 (2004).
  
\bibitem{fischerquark} C.\,S.~Fischer and M.\,R.~Pennington,
  Phys.\ Rev.\ \textbf{D\,73}, 034029 (2006).
  
\bibitem{sauli} V.~\v{S}auli, Few Body Syst.\ \textbf{39}, 45 (2006).

\bibitem{bhagwatpanic} M.\,S.~Bhagwat and P.\,C.~Tandy, 
  AIP Conf.\ Proc.\  {\bf 842}, 225 (2006).
    
\bibitem{nickel} D.~Nickel, ``Extraction of Spectral Functions from Dyson-Schwinger Studies via the Maximum Entropy Method,'' hep-ph/0607224.

\bibitem{cdragw} C.\,D.~Roberts and A.\,G.~Williams,
  Prog.\ Part.\ Nucl.\ Phys.\  {\bf 33}, 477 (1994).
    
\bibitem{Roberts:2000aa}
C.\,D.~Roberts and S.\,M.~Schmidt, Prog.\ Part.\ Nucl.\ Phys.\ \textbf{45}, S1 (2000).

\bibitem{ralvs} R.~Alkofer and L.~von Smekal,
  Phys.\ Rept.\ \textbf{353}, 281 (2001).
  
\bibitem{pmcdr} P.~Maris and C.\,D.~Roberts,
  Int.\ J.\ Mod.\ Phys.\ \textbf{E\,12}, 297 (2003).
  
\bibitem{cf06} C.\,S.~Fischer,
  J.\ Phys.\ {\bf G\,32}, R253 (2006).
  
\bibitem{heavyheavy} M.~S.~Bhagwat, A.~Krassnigg, P.~Maris and C.~D.~Roberts, ``Mind the gap,'' nucl-th/0612027.

\bibitem{bhagwatmaris} M.~S.~Bhagwat and P.~Maris, ``Vector meson form factors and their quark-mass dependence,'' nucl-th/0612069.

\bibitem{Holl:2004fr}
A.~H\"oll, A.~Krassnigg, and C.\,D.~Roberts, Phys.\ Rev.\ {\bf C\,70}, 042203(R) (2004).

\bibitem{arnebeijing} A.~H\"oll, A.~Krassnigg, C.~D.~Roberts and S.~V.~Wright,
  Int.\ J.\ Mod.\ Phys.\ {\bf A\,20}, 1778 (2005).
  
\bibitem{andreasradialgg} A.~H\"oll, A.~Krassnigg, P.~Maris, C.\,D.~Roberts and S.\,V.~Wright,
  Phys.\ Rev.\ {\bf C\,71}, 065204 (2005).
  
\bibitem{Krassnigg:2006ps} A.~Krassnigg, C.\,D.~Roberts and S.~V.~Wright, ``Meson spectroscopy and properties using Dyson-Schwinger equations,'' nucl-th/0608039.
  
\bibitem{a1b1} J.\,C.\,R.~Bloch, Yu.\,L.~Kalinovsky, C.\,D.~Roberts and S.\,M.~Schmidt, 
Phys.\ Rev.\ \textbf{D\,60}, 111502 (1999).

\bibitem{burdenexotic} C.\,J.~Burden and M.\,A.~Pichowsky,
  Few Body Syst.\ {\bf 32}, 119 (2002).

\bibitem{watsona1} P.~Watson, W.~Cassing and P.\,C.~Tandy,
  Few Body Syst.\  {\bf 35}, 129 (2004).
  
\bibitem{krassnigga1}  A.~Krassnigg,
  Phys.\ Rev.\ {\bf C\,72}, 028201 (2005).
  
\bibitem{arneN} R.~Alkofer, A.~H\"oll, M.~Kloker, A.~Krassnigg and C.\,D.~Roberts,
  Few Body Syst.\  {\bf 37}, 1 (2005).

\bibitem{arneN2} A.~H\"oll, R.~Alkofer, M.~Kloker, A.~Krassnigg, C.\,D.~Roberts and S.\,V.~Wright,
  Nucl.\ Phys.\ A {\bf 755}, 298 (2005).

\bibitem{Flambaum:2005kc}
  V.\,V.~Flambaum, A.~H\"oll, P.~Jaikumar, C.\,D.~Roberts and S.\,V.~Wright,
  Few Body Syst.\  {\bf 38}, 31 (2006).

\bibitem{arneN3} M.\,S.~Bhagwat, A.~H\"oll, A.~Krassnigg and C.\,D.~Roberts,
``Theory and phenomenology of hadrons,'' nucl-th/0610080.

\bibitem{bhagwatpoles} M.~Bhagwat, M.\,A.~Pichowsky and P.\,C.~Tandy,
  Phys.\ Rev.\ {\bf D\,67}, 054019 (2003).

\bibitem{latticemem} M.~Asakawa, T.~Hatsuda and Y.~Nakahara,
  Prog.\ Part.\ Nucl.\ Phys.\  {\bf 46}, 459 (2001).

\bibitem{Leinweber:2004it} D.\,B.~Leinweber, W.~Melnitchouk, D.\,G.~Richards, A.\,G.~Williams and J.\,M.~Zanotti,
  Lect.\ Notes Phys.\  {\bf 663}, 71 (2005).

\bibitem{Michael:1985ne} C.~Michael,
  Nucl.\ Phys.\ {\bf B\,259}, 58 (1985).

\bibitem{Luscher:1990ck} M.~L\"uscher and U.~Wolff,
  Nucl.\ Phys.\ {\bf B\,339}, 222 (1990).

\bibitem{Maris:1997hd}
P.~Maris, C.\,D.~Roberts and P.\,C.~Tandy, Phys.\ Lett.\ \textbf{B\,420}, 267 (1998).

\bibitem{Maris:1997tm}
P.~Maris and C.\,D.~Roberts, Phys.\ Rev.\  \textbf{C\,56}, 3369 (1997).

\bibitem{changlei} L.~Chang, Y.-X.~Liu, M.\,S.~Bhagwat, C.\,D.~Roberts and S.\,V.~Wright, Phys.\ Rev.\ C \textbf{75}, 015201 (2007).

\bibitem{maristandy1} P.\ Maris and P.\,C.\ Tandy,
Phys.\ Rev.\ \textbf{C\,60}, 055214 (1999).

\bibitem{raya} P.~Maris, A.~Raya, C.\,D.~Roberts and S.\,M.~Schmidt,
  Eur.\ Phys.\ J.\ A {\bf 18}, 231 (2003).

\bibitem{llewellyn} C.\,H.~Llewellyn-Smith,
  Annals Phys.\  {\bf 53}, 521 (1969).

\bibitem{Truong:1988zp} T.\,N.~Truong, Phys.\ Rev.\ Lett.\ \textbf{61}, 2526 (1988);
\textit{ibid}.\ \textbf{67}, 2260 (1991).

\bibitem{Dobado:1992ha} A.~Dobado and J.\,R.~Pelaez, Phys.\ Rev.\ \textbf{D\,47}, 4883 (1993) 4883.

\bibitem{Hoell:2005st}
A.~H{\"o}ll, P.~Maris, C.\,D.~Roberts, and S.\,V.~Wright, ``Schwinger functions
  and light-quark bound states, and sigma terms,'' nucl-th/0512048.

\bibitem{mcneile} C.~McNeile, ``Meson and baryon spectroscopy on a lattice,''
in \textit{Int.\ Rev.\ Nucl.\ Phys., Vol.\ 9}, ed. A.\,M.~Green (World Scientific, Singapore, 2004) pp.~1-101.

\bibitem{richards} D.\,G.~Richards, ``Lattice gauge theory: QCD from quarks to hadrons,'' nucl-th/0006020.

\bibitem{Basak:2004hr} S.~Basak {\it et al.}  [LHP Collaboration],
  Nucl.\ Phys.\ Proc.\ Suppl.\  {\bf 140}, 278 (2005).

\bibitem{Morningstar:2005pv} C.~Morningstar,
  Int.\ J.\ Mod.\ Phys.\ {\bf A\,21}, 843 (2006).
  
\bibitem{Juge:2006gr} K.~J.~Juge {\it et al.}, ``Towards a determination of the spectrum of QCD using a space-time lattice,'' hep-lat/0601029.

\bibitem{Graz} T.~Burch, C.~Gattringer, L.~Y.~Glozman, C.~Hagen, C.~B.~Lang and A.~Schafer,
  Phys.\ Rev.\ {\bf D\,73}, 094505 (2006);
T.~Burch, C.~Gattringer, L.~Y.~Glozman, C.~Hagen, D.~Hierl, C.~B.~Lang and A.~Schafer,
  Phys.\ Rev.\ D {\bf 74}, 014504 (2006).

\bibitem{Anisovich:2000kx}
A.~V. Anisovich, V.~V. Anisovich and A.~V. Sarantsev, Phys.\ Rev.\ \textbf{D\,62}, 051502  (2000).

\end{thebibliography}
\end{document}